\documentclass[pdflatex,sn-nature]{sn-jnl}
\usepackage{graphicx} 
\usepackage[version=3]{mhchem} 
\usepackage{xcolor}
\usepackage{soul} 
\usepackage{array}
\usepackage{graphicx}%
\usepackage{multirow}%
\usepackage{amsmath,amssymb,amsfonts}%
\usepackage{amsthm}%
\usepackage{mathrsfs}%
\usepackage[title]{appendix}%
\usepackage{xcolor}%
\usepackage{textcomp}%
\usepackage{manyfoot}%
\usepackage{booktabs}%
\usepackage{algorithm}%
\usepackage{algorithmicx}%
\usepackage{algpseudocode}%
\usepackage{listings}%

\title{Bridging Molecular Scales with Implicit Score Matching for Bottom-Up Coarse Graining}

\author*[1]{Patrick G. Sahrmann}
\email{sahrmann@lanl.gov}
\author[2]{Nicholas Lubbers}
\author[1]{Benjamin T. Nebgen}
\author[2]{Emily Shinkle}
\author[1,3]{Sergei Tretiak}
\author[1]{Kipton Barros}
\author[1]{Brenden W. Hamilton}

\affil[1]{Theoretical Division, Los Alamos National Laboratory, Los Alamos, New Mexico 87545, USA}
\affil[2]{Computing and Artificial Intelligence Division, Los Alamos National Laboratory, Los Alamos, New Mexico 87545, USA}
\affil[3]{Center for Integrated Nanotechnologies, Los Alamos National Laboratory, Los Alamos, New Mexico 87545, USA}

\date{September 2026}

\begin{document}

\maketitle

\enlargethispage{2\baselineskip}

\begin{abstract}

Molecular dynamics simulations provide a computational microscope for atomic scale processes but remain restricted to relatively small spatial and temporal scales. 
Coarse-grained models extend their reach by representing groups of atoms as effective interaction sites. 
The number of particles is systematically reduced by embedding the collective behavior of groups of atoms into coarse-grained sites governed by effective potentials.
However, constructing accurate, highly coarsened potentials, even with the assistance of modern machine-learning methods, is a grand challenge because coarse grained forces depend on a statistical distribution of composite atomic configurations making direct calculation computationally prohibitive. 
Here, we introduce an alternative framework based on implicit score matching that directly avoids the need for explicit coarse-grained forces. 
This approach both substantially reduces the training data required relative to conventional force-matching methods and enables high-fidelity models in which each coarse-grained site represents hundreds to thousands of atoms. 
We demonstrate the method's versatility and effectiveness across a diverse range of chemical systems and phenomena, developing bottom-up, machine-learned models. These enable efficient modeling at micrometer and millimeter length scales, 
while retaining significant amounts of the atomic scale fidelity.
These results establish implicit score matching as a practical route towards chemically accurate simulations that bridge molecular and mesoscopic scales.


    
\end{abstract}

\newpage

\section{Introduction}

Over the past decade, machine learning (ML) has transformed computational chemistry and computational materials science, enabling predictive modeling at unprecedented accuracy and scale. 
In particular, ML models can accurately represent the potential-energy surfaces of atomic and molecular systems.\cite{keith2021combining,behler2016perspective,meuwly2021machine,kulichenko2024data,deringer2019machine} 
Replacing computationally demanding quantum-chemical calculations with ML potentials has emerged as one of the most promising applications of ML in chemistry.\cite{kulichenko2023AL,chen2022universal,griesemer2023accelerating,li2022deep,smith2018less,smith2019approaching,zhang2024exploring,batatia2022mace,hansen2015machine,chen2025mean,hamilton2023high} 
These models enable molecular dynamics (MD) simulations with near-quantum accuracy at spatiotemporal scales several orders of magnitude beyond those accessible to direct quantum-mechanical simulations. 
A remaining grand challenge is further extending this advance to the mesoscopic and macroscopic scales without sacrificing molecular-level accuracy.\cite{praprotnik2008scale}


Coarse-graining (CG) offers a promising route towards bridging these scales by reducing the number of degrees of freedom, namely, grouping several atoms to represent them collectively as effective meso-particles.\cite{jin2022CG,noid2023perspective,marrink2023cg} 
This strategy has enabled the development of numerous high-fidelity, ML CG force fields in which each site captures the collective behavior of up to several dozen atoms per CG site.\cite{wang2019cg,wang2019acg,chan2019cg,husic2020cg,majewski2023machine,lee2024rdx,duschatko2024al,campos2024machine,charron2025navigating,li2025coarse}
Reaching substantially larger scales, however, requires more aggressive CG modeling and correspondingly more efficient and judicious approaches to model construction. 


Bottom-up CG models are typically trained using high-fidelity atomistic simulations as reference data, akin to how atomistic force fields are trained against quantum-chemical calculations.\cite{jin2022CG,noid2023perspective,durumeric2023ml,sahrmann2025ml}
Unlike an atomistic potential-energy surface, however, neither the effective CG potential nor its forces can be obtained directly from an atomistic trajectory because they incorporate the averaged effects of the eliminated degrees of freedom. Thus, evaluating this free-energy surface requires calculation of high-dimensional constrained ensemble averages. 
Obtaining well-converged, ground truth forces, particularly through constrained MD, can be computationally prohibitive.\cite{duschatko2024al,chen2025mean,park2026cg}
Most machine-learned CG force fields are therefore developed using force-matching (FM) in which the forces predicted for the CG sites are fitted to mapped atomistic forces.\cite{noid2008MSCG,wang2019cg} 
These mapped forces provide noisy estimates of the ground truth CG forces which renders the FM method extremely data-inefficient.


To truly bridge the scales between atomistic and mesoscopic regimes, CG sites should represent hundreds to thousands of atoms. 
However, as a CG site encompasses increasingly larger groups of atoms, the corresponding force estimates become progressively more variable because a much larger ensemble of atomistic configurations maps onto a single CG configuration. This increased variance substantially complicates the construction of accurate CG force fields.
Additionally, the use of non-linear mappings of atoms to CG sites, which are pivotal in the extreme CG regime, introduces Jacobian terms to the CG force which are computationally prohibitive for large-scale systems.\cite{kalligiannaki2015force,pak2019VCG}
Distribution-based methods such as relative entropy minimization avoid these issues, however this comes at the cost of long simulations of the CG model wherein the potential is iteratively updated and may become numerically unstable.\cite{shell2008entropy,thaler2022cg,sahrmann2024cg}
We introduce in this work a method for training reliably high-fidelity ML models without needing expensive force calculations, allowing for truly multiscale modeling. 

In this Article, we develop the implicit score-matching (ISM) method that bypasses these issues by recasting the minimization of the force residual as a minimization condition between the expectation of the CG model force and the CG model Hessian.\cite{Hyvarinen2005score} 
The ISM method avoids fitting to noisy forces during optimization, suggesting the possibility of training CG potentials with far greater data efficiency.
Furthermore, by removing the need to calculate the force on the CG site directly, the computational cost of building an ML model where each site represents thousands of atoms immediately becomes tractable as prohibitive calculations of the Jacobian component of the CG force are avoided entirely.
This enables a far more diverse range of possible CG representations which are much coarser than commonly employed center-of-mass or implicit solvent representations.

The ISM method thus enables the construction of CG models with unprecedented accuracy and that are capable of accessing previously unreachable physical scales through extreme atom aggregation in defining CG sites. Using a set of diverse chemical examples, we demonstrate 1) how the ISM method can be used to dramatically improve the data-efficiency in training ML CG potentials when forces are available,
2) ISM's unique strengths, as a standalone method, in developing CG potentials at coarse resolutions far beyond what conventional ML CG methods have considered, 
and 3) how such ML CG models can capture physics at the micrometer and even millimeter scales rigorously from the bottom-up, atomistic-level physics.


\section{Results}

\subsection{Learning from Implicit Scores}


The fundamental goal of CG modeling is to construct a CG potential $U_\theta(\mathbf{R})$ with tunable parameters $\theta$ which accurately captures the behavior of the atomistic system mapped to the CG resolution.
Here, we assume a collection of $n$ atomic positions $\mathbf{r}$ with microscopic potential energy $u(\mathbf{r})$.
The CG potential itself is a constrained partition function which is analytically intractable, and hence the majority of efforts in learning the CG potential have focused on force-based training.
Hence, employing the atomistic forces directly rather than the ground truth forces in training the CG model as in FM is a significantly computationally cheaper alternative.

For the coarsening mapping operation $\boldsymbol{\mathcal{M}}(\mathbf{r}) = \mathbf{R}$, the forces of the CG potential can be regressed onto the atomistic forces, $\mathbf{f}_i(\mathbf{r}) = -\nabla_i u(\mathbf{r})$, in the following manner to produce the FM loss function,\cite{noid2008MSCG,kalligiannaki2015force}
\begin{equation}\label{Eq. 1}
    \mathcal{L}_{FM}(\theta)  = \frac{1}{3N} \sum_{I=1}^N\mathbb{E}_{\mathbf{r} \sim p_{AA}} [ \lVert \boldsymbol{\Xi}_I (\mathbf{f}(\mathbf{r}) )+ \nabla_I  U_\theta(\boldsymbol{\mathcal{M}}(\mathbf{r}))\rVert^2 ],
\end{equation}
where $\Xi$ represents the force-mapping operator, $p_{AA}$ is the configurational distribution of the reference all-atom (AA) system, and $I$ indexes the $N$ coarse-grained particles.
Importantly, calculation of the force-mapping operation necessitates the inclusion of Jacobian terms when the spatial mapping is non-linear, which becomes prohibitively expensive for large-scale systems, 
such that the majority of CG mapping operations considered in the literature are restricted to linear, static operations.

We propose to bypass these complications of employing FM via ISM. 
It can be shown (see Supplementary Section 1) that the ISM loss function shares the same minimum as FM and can be expressed as
\begin{equation}\label{Eq. 2}
    \mathcal{L}_{ISM}(\theta)  = \frac{1}{3N}\Big [\sum_{I=1}^N\mathbb{E}_{\mathbf{r} \sim p_{AA}} [\lVert \nabla_I U_\theta(\boldsymbol{\mathcal{M}}(\mathbf{r}))\rVert^2]  - \frac{2}{\beta} \cdot \sum_{I=1}^N \mathbb{E}_{\mathbf{r} \sim p_{AA}}[\nabla^2_I U_\theta (\boldsymbol{\mathcal{M}}(\mathbf{r}))]\Big ],
\end{equation}
 where $\beta = 1/k_BT$. That is, matching the forces of the CG model and ground truth CG potential is mathematically equivalent to minimizing the difference in the expectation of the square of the CG model force and its temperature-weighted Hessian.\cite{hutchinson1990estimator,song2020score}
Crucially, the ISM loss function  does not necessitate the calculation of either the ground truth or mapped atomistic forces at all during optimization.
Consequently, there are two key potential advantages of the ISM loss function over FM:
1) ISM avoids fitting to noisy forces, potentially conferring improved data efficiency,
2) complications with utilizing non-linear mappings are eliminated when conducting ISM by eliminating the need for the prohibitively expensive Jacobian calculation. 
The remainder of this article is spent showcasing the inherent power and capabilities of the ISM method across different chemical problems.

\subsection{Application: Buckminsterfullerene}

\begin{figure}[htpb]
        \centering
        \includegraphics[width=0.7\linewidth]{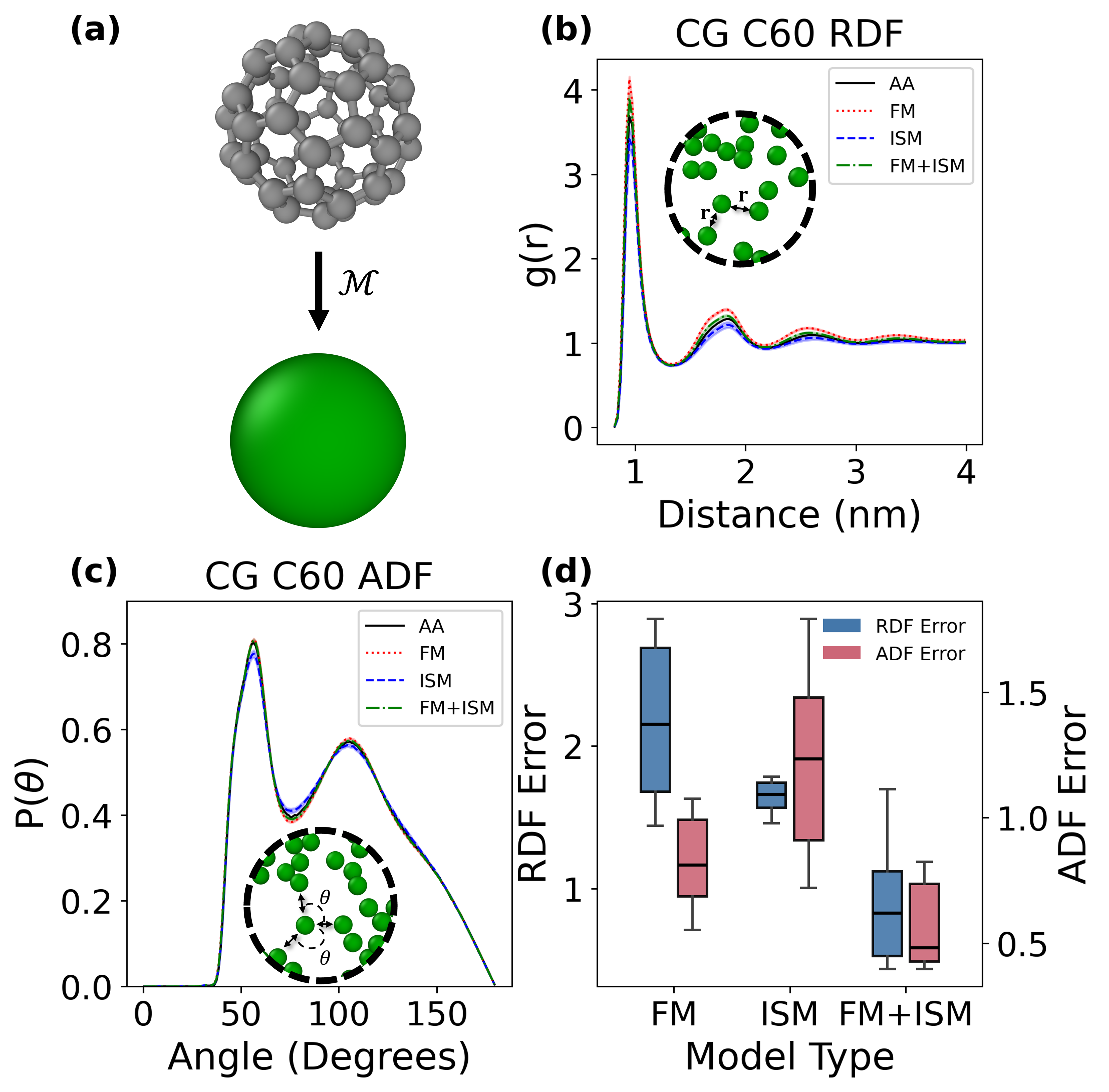}
        \caption{(a) Mapping of $\text{C}_{60}$ to its center-of-mass is shown. (b) RDFs of all models are shown. Standard errors are shown for $n = 10$ models. Graphical portrayal of radial calculations are shown (inset). (c) ADFs of all models including standard error are shown. Graphical portrayal of angular calculations are shown (inset). (d) Box plot is shown of the ensemble model performance amongst the model classes considered.}
        \label{fig:c60}
\end{figure}

Firstly, we demonstrate the utility of ISM in learning ML CG potentials when forces are available for FM.
We consider a linear center-of-mass mapping of buckminsterfullerene, or carbon buckyballs ($\text{C}_{60}$), as shown in Fig. \ref{fig:c60}(a), in the condensed phase at $T = 2000$ K.
At this temperature, $\text{C}_{60}$ behaves as a liquid-vapor mixture with variations in local density which is difficult for CG models to capture.

We construct three classes of models trained from FM, ISM, and a hybrid mixture of the two minimization conditions (FM+ISM). 
Force-based validation metrics are not strictly correlated with the simulation behavior (see Supplementary Table S3), as observed in previous findings.\cite{fu2023forces}
We consider two metrics for model error, namely the radial distribution function (RDF) and angular distribution function (ADF) within the first solvation shell (see Supplementary Section 3).
We plot the structural behavior of the developed models according to these metrics in Figures \ref{fig:c60}(b) and \ref{fig:c60}(c), respectively.

From this, we see the FM models have a strong tendency to bias towards the liquid phase, whereas ISM models slightly bias towards the vapor phase.
Intriguingly, the FM+ISM models correctly balance this phase-coexistence to produce the correct bulk structure. 
It is apparent that the ISM model, in general, captures 2-body correlations better than the FM models.
As shown in Fig. \ref{fig:c60}(c), FM  and FM+ISM models both capture 3-body correlations better than ISM models. 
Remarkably, we find that employing both FM and ISM as losses in training substantially reduces the error for both metrics in the resulting models, as exhibited in Fig.  \ref{fig:c60}(d).
Furthermore, FM+ISM models still outperform either FM or ISM models when an order of magnitude less training data is employed (see Supplementary Fig. S1).
These findings suggest that ISM can be used not only for the development of more accurate CG models but can be used in conjunction with other loss functions to dramatically reduce the amount of training data required.

\subsection{Application: Mesoscale Coarse-Graining of Water}


A key advantage of ISM is its flexibility to define each CG site as a non-linear averaging of the atomic-scale variables. 
This differs sharply from conventional FM, which relies crucially on linearity for its simple training protocol. 
This linearity imposes strict limitations on the degree of coarsening that can be explored for CG modeling, i.e., a linear mapping cannot suitably describe how multiple molecules can be mapped to a CG site.

As a first example of a nonlinear-mapped CG model, we consider water.
The representation of a fluid element wherein multiple molecules constitute a single CG site is a common means of accessing mesoscale fluid dynamics, such as in dissipative particle dynamics.\cite{español1995}
The central difficulty in learning ML CG potentials here is that the static nature of the mapping operator required for FM learning is fundamentally incompatible with mesoscale mappings which require dynamic assignment of molecules to CG sites. \cite{han2018cg}

\begin{figure}[htpb]
        \centering
        \includegraphics[width=0.4\linewidth]{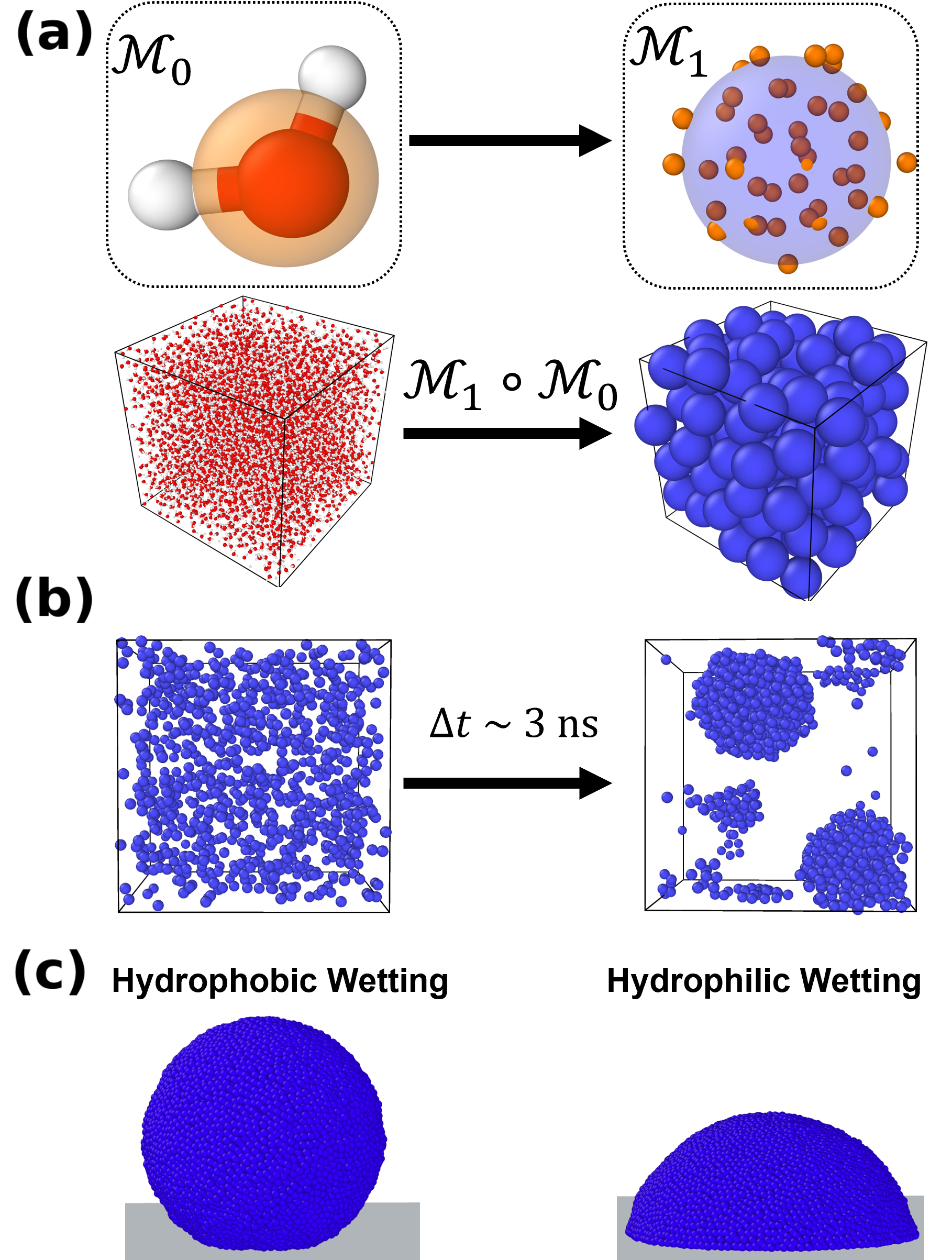}
        \caption{(a) Mesoscale mapping of water is conducted in a two step process. First, individual water molecules are mapped to their center of mass (upper left). After this mapping is conducted, a soft Voronoi tessellation is performed which assigns water molecules to a CG blob (upper right). The full mapping operation is summarized (bottom). (b) The resulting CG blob model is capable of spontaneous droplet formation in vacuum. (c) The CG water models developed are capable of surface wetting on both hydrophobic and hydrophilic model surfaces.}
        \label{fig:water_map}
\end{figure}

We consider the CG modeling of water such that water `blobs' are defined from the atomistic simulation of water.
Our mapping operation is equivalent to a soft Voronoi tessellation which must be solved self-consistently at every CG configuration (see Supplementary Section 3.2).
The local density of each molecular center-of-mass point around a CG reference point is used to inform the position of the next CG reference point iteratively within the loop.
This is repeated until the reference point does not change in position up to a given tolerance,
see Fig. \ref{fig:water_map}(a).
We find that this self-consistent mapping is robust across varying initial conditions and dynamically assigns molecules to mesoscopic CG sites in a spatially homogeneous manner (see Supplementary Figs. S2 and S3).
To compare to an FM based model, due to the difficulty in extracting exact mapped forces for this mapping, we employ the blob momentum time derivative as a proxy for explicit forces.
We consider a range of degrees of coarsening utilizing this mapping which we denote by $n/N$ where $n$ is the number of water molecules and $N$ is the number of CG sites at that resolution.

\begin{figure}[h]
        \centering
        \includegraphics[width=0.8\linewidth]{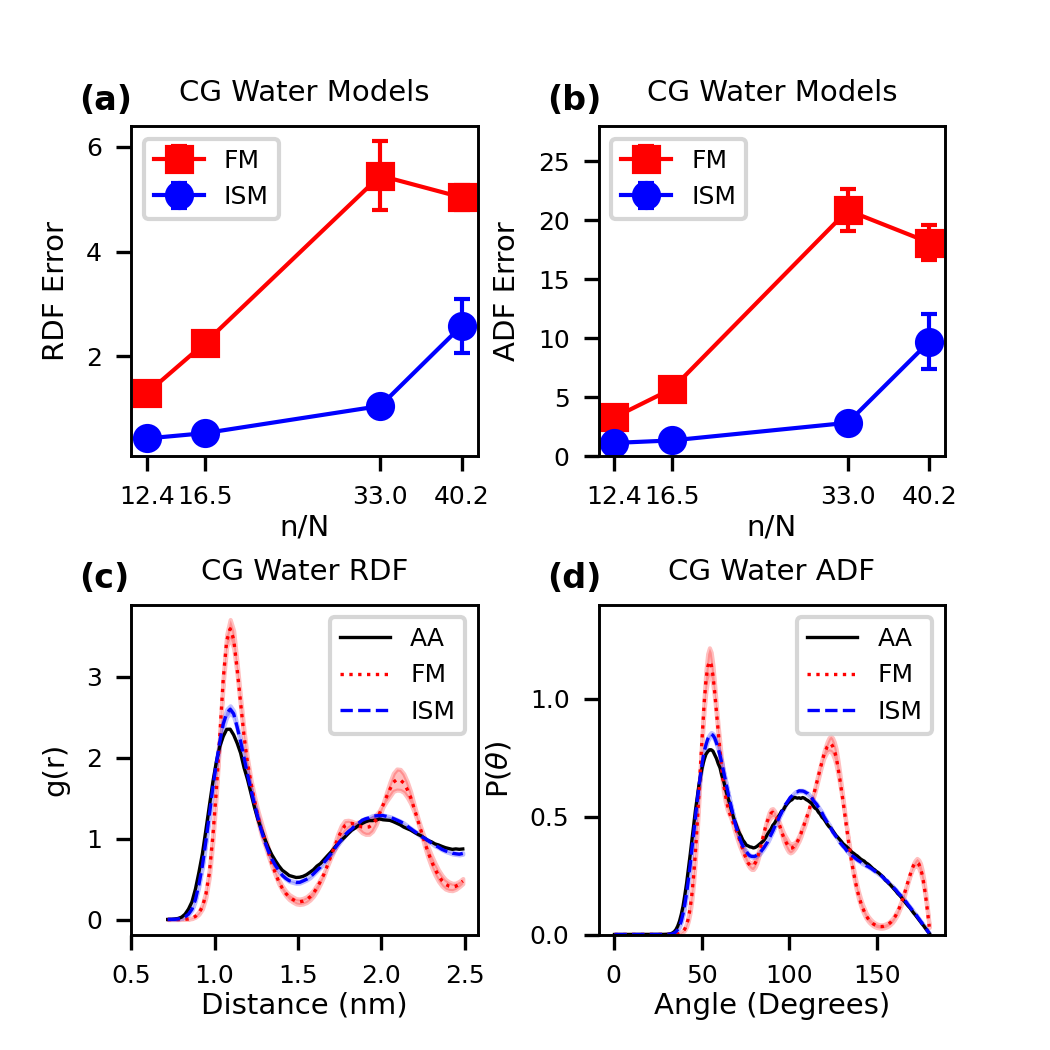}
        \caption{(a) RDF and (b) ADF errors of water are shown for different degrees of coarsening. (c) The RDF and (d) ADF of the n/N = 33 CG water models are shown. Standard errors are shown for $n = 8$ models.}
        \label{fig:water_results}
\end{figure}

We employ the same error metrics as in the previous example to assess model performance.
We find that ISM CG models consistently outperform FM CG models in terms of RDF and ADF errors across differing degrees of coarsening, as demonstrated in Figs. \ref{fig:water_results}(a) and \ref{fig:water_results}(b), respectively.
Additional RDF and ADF results are provided in Supplementary Figs. S4 and S5, respectively.
Furthermore, the ISM models developed recapitulate various collective phenomena of water, including spontaneous droplet formation, as shown in Fig. \ref{fig:water_map}(b), under low density and both hydrophobic and hydrophilic wetting, as shown in Fig. \ref{fig:water_map}(c). 
Additional results for CG models trained from interfacial water, including distribution functions and droplet shape and contact angle behavior, are provided in Supplementary Figs. S6 and S7, respectively.
This suggests that, despite the extremely coarse representation of water employed in this work, which has completely removed individual molecular identity, ISM based potentials can capture emergent liquid behavior.
Furthermore, such models are able to capture fluid behavior in a transferable and predictive manner, as spontaneous droplet formation and wetting behaviors were not included in the training data. 
In Supplementary Section 3.3, we discuss how material properties such as the isothermal compressibility (see Supplementary Fig. S8) can be extracted from these CG simulations and the obtained correlation functions, bridging extremely coarse representations to bulk thermodynamic behavior. 
Thus, by eliminating the prohibitive force calculations that have constrained bottom-up mesoscale modeling, the ISM framework enables accurate descriptions of fluids at exceptionally coarse resolutions.

\subsection{Application: Mesoscale Coarse-Graining of a Lipid Bilayer}

Most biophysical phenomena regarding the cell membrane occur on scales that are out of reach for atomistic MD. 
Both bottom-up ML CG models of lipid bilayers as well as empirically developed force-fields have focused on CG operations which retain individual lipid molecular identity. \cite{souza2021martini,sahrmann2024cg,sahrmann2025cg,liao2025cg,majumder2025cg} 
These models enable observation of, e.g., local membrane curvature and domain formation, but are only computationally practical for scales of tens to hundreds of nanometers.

ISM readily enables non-linear mappings, and hence we develop a mapping operation in which a single CG site corresponds to an entire patch of a lipid bilayer and train a model to rigorously capture micron-scale membrane physics from the bottom-up.

We consider the mesoscale CG modeling of a 1,2-dioleoyl-\emph{sn}-glycero-3-phosphocholine (DOPC) lipid bilayer, a prototypical model system for cellular membranes.
We employ a non-linear CG mapping operation which probes the local density of lipid groups around a mesh grid which comprise the midplane (see Supplementary Section 3.3).

Fig. \ref{fig:DOPC_map} provides quantification of the reduction in particle number enabled by this non-linear mapping, where we observe that our mapping operation reduces the particle number by a factor of $10^4$.
This extreme CG mapping enables the effective representation of billions of atoms, at the frontier of modern atomistic MD capabilities, within tractable simulations.
Our lipid bilayer simulation here is the all-atom equivalent of simulating $\approx1.4 \times 10^9$ atoms.

\begin{figure}[htpb]
        \centering
        \includegraphics[width=0.45\linewidth]{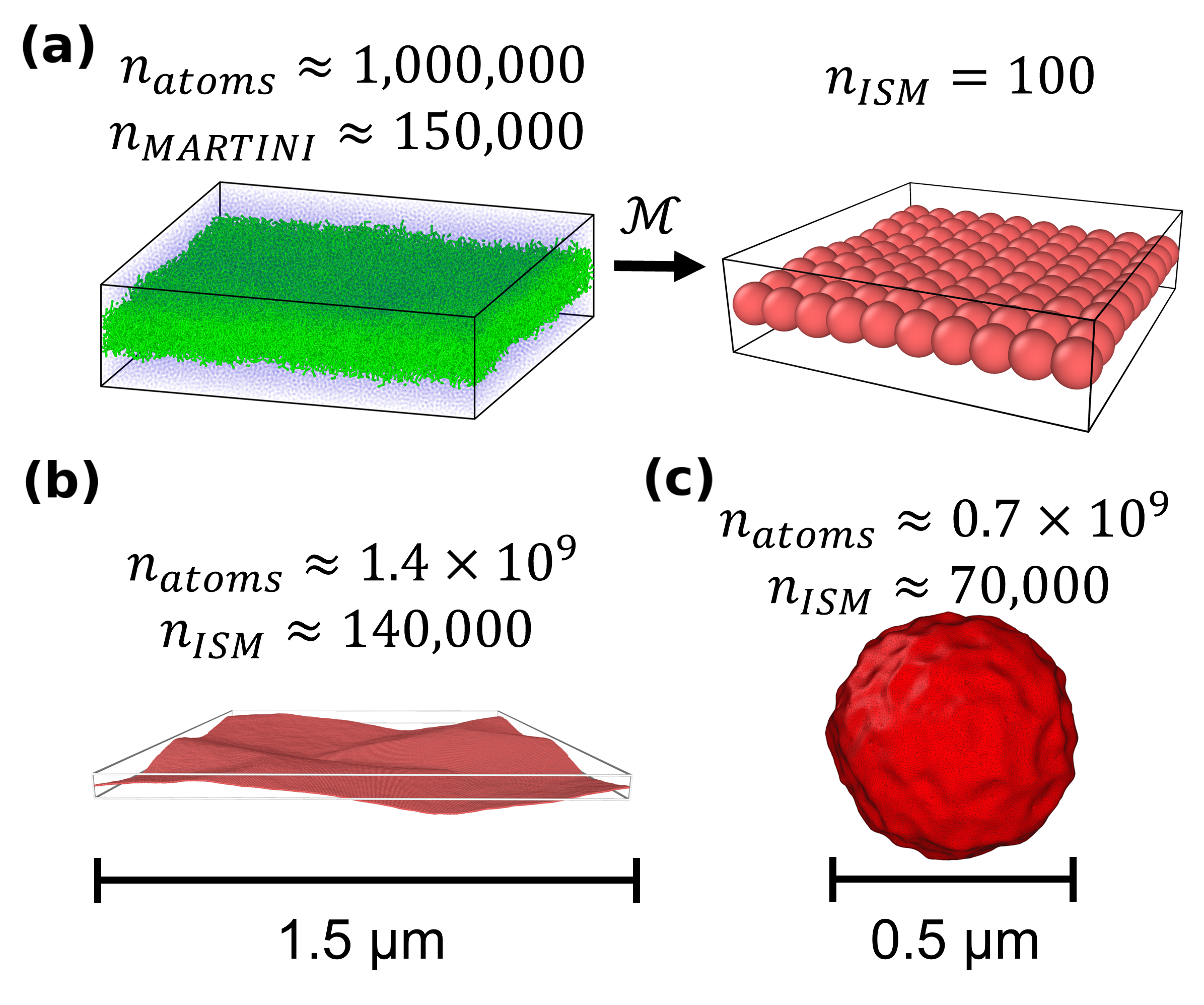}
        \caption{(a) CG mapping is shown from a Martini CG patch to the mesoscale CG mapping considered in this work. (b) Large-scale membrane and (c) vesicle simulations of the ISM CG models are shown. For all simulations, comparisons of particle number are provided amongst the models considered.}
        \label{fig:DOPC_map}
\end{figure}

\begin{figure}[htpb]
        \centering
        \includegraphics[width=1\linewidth]{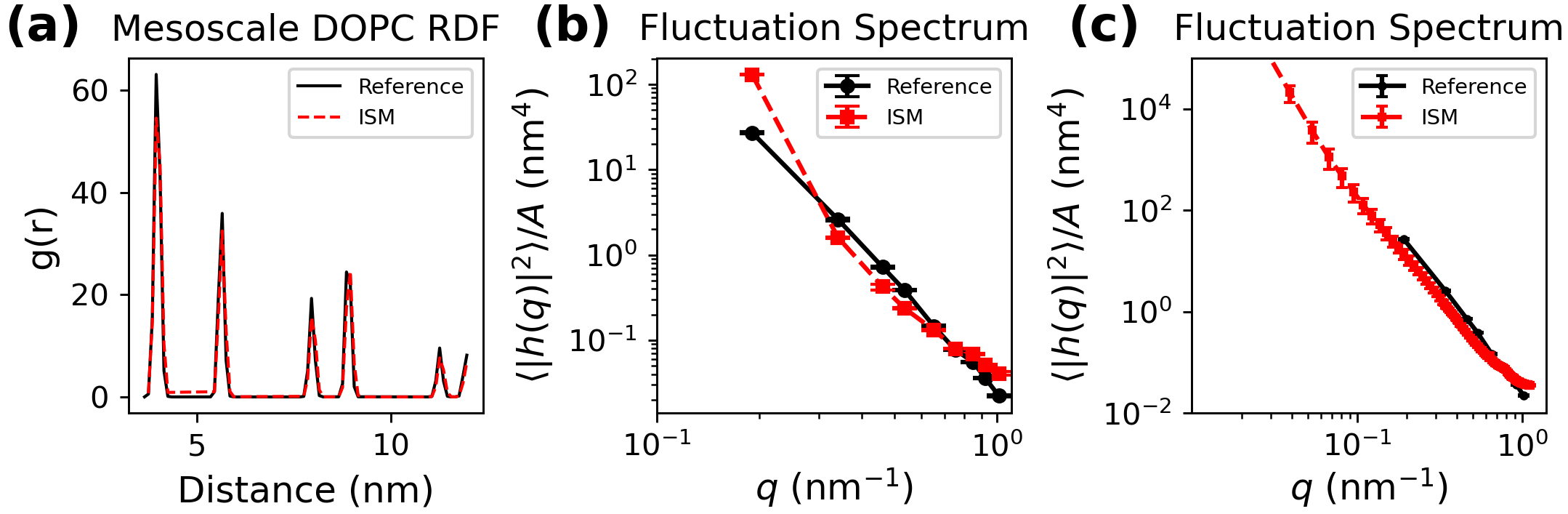}
        \caption{(a) RDF is shown for the reference and ISM models. (b) Fluctuation spectrum for the reference and ISM models are shown for lipid membranes whose sizes equal the reference model simulation dimensions. (c) Comparison of the fluctuation spectrum for a 1.5 $\mu$m $\times$ 1.5 $\mu$m ISM CG patch against the reference model.}
        \label{fig:DOPC_results}
\end{figure}

Quantitative analysis of the lipid patch RDF shows good agreement with the mapped atomistic model, as shown in Fig. \ref{fig:DOPC_results}.
It has been observed that, in the long wavelength regime, membranes across a wide range of chemical complexity are well-described by the Canham-Helfrich continuum model.\cite{helfrich1973,baumgart20007lipid,dimova2014lipid,sadeghi2020cg}
A fundamental result of Canham-Helfrich theory is that, under zero surface tension, the membrane fluctuation spectrum, $h$, is connected to the wave number, $q$, via a power-law. 
We find that the ISM model follows this trend in the long wavelength limit where such wavelengths are not reachable by atomistic or conventional CG MD.
We report bending moduli values of $\kappa $ = 34.34 $k_BT$ and $\kappa $ = 51.35 $k_BT$ for the mapped Martini model and the ISM model, respectively (see Supplementary Fig. S9).  
Atomistic and experimental values of the bending modulus for DOPC have been reported as 28.8 and 18.3 $k_BT$, respectively, which produce notably softer membranes than the ISM model.\cite{venable2015lipid,pan2008lipid}
This behavior has been observed repeatedly for other bottom-up CG lipid models, and can be attributed to the difficulty in capturing the many-body behavior which encapsulates membrane bending.\cite{pak2019VCG,sahrmann2024cg}
However, we observe that our ISM model produces a better fit to Canham-Helfrich theory, i.e., a $q^{-4}$ dependence, than the mapped Martini model.
This is likely due to two factors, one being that the ISM model is able to access much longer wavelengths, for which Canham-Helfrich theory is valid, and the other being that our CG mapping to the mid-plane is essentially an element discretization of Canham-Helfrich theory.

The amphipathic nature of lipids enables their emergent assembly into a diverse array of morphologies under variations in concentration.
The ISM CG model displays predictive capability in morphological behavior, including vesicle stability (see Fig. \ref{fig:DOPC_map}(c)). 
Given that the ISM CG model produces both stable bilayer and vesicle morphologies, it can be concluded that the amphipathic forces which drive shape formation are largely retained even at mesoscale degrees of coarsening.
We furthermore find that the ISM CG model reproduces the behavior of lipid bilayers to buckle under stress (see Supplementary Fig. S9),
which provides further evidence that the biomechanics is largely retained by the ISM model and demonstrates that this extremely coarse model possesses predictive capability.

\subsection{Application: Macroscale Coarse-Graining}

Coarsening from the atomistic to macroscale must, as a matter of practicality, proceed in a hierarchical fashion.
At each level of coarse-graining resolution, there exists a characteristic spatial scale beyond which generating the training data required for model construction becomes computationally impractical due to the prohibitively large number of particles involved. 
This necessitates, first, the development of high-fidelity CG models such that physics can be accurately transferred at each scale, and second, the employment of non-linear mappings to coarsen beyond the molecular level.
We have shown repeatedly thus far that the ISM method, in conjunction with ML modeling, is uniquely poised to address these problems by enabling development of highly accurate, non-linearly mapped CG models.

We consider successive CG modeling of the mesoscale water model developed in this work.
The iterative mapping operation consists of large-scale simulations of the CG model at the current iteration followed by the fixed point mapping operator previously employed in the mesoscale CG water example.
Autonomous development of a hierarchy of CG models utilizing this approach presents practical challenges in ensuring the configuration space, which grows exponentially in size with each iteration, is adequately sampled and stable training procedures are maintained.
To resolve this issue, the models trained at each iteration do not receive bare coordinate values as input but instead are fed rescaled coordinates which preserve correlation lengths at each iteration, significantly stabilizing training.
This operation is closely related to Wilsonian renormalization (see Supplementary Section 3.4).\cite{wilson1971}

\begin{figure}[htpb]
        \centering
        \includegraphics[width=0.8\linewidth]{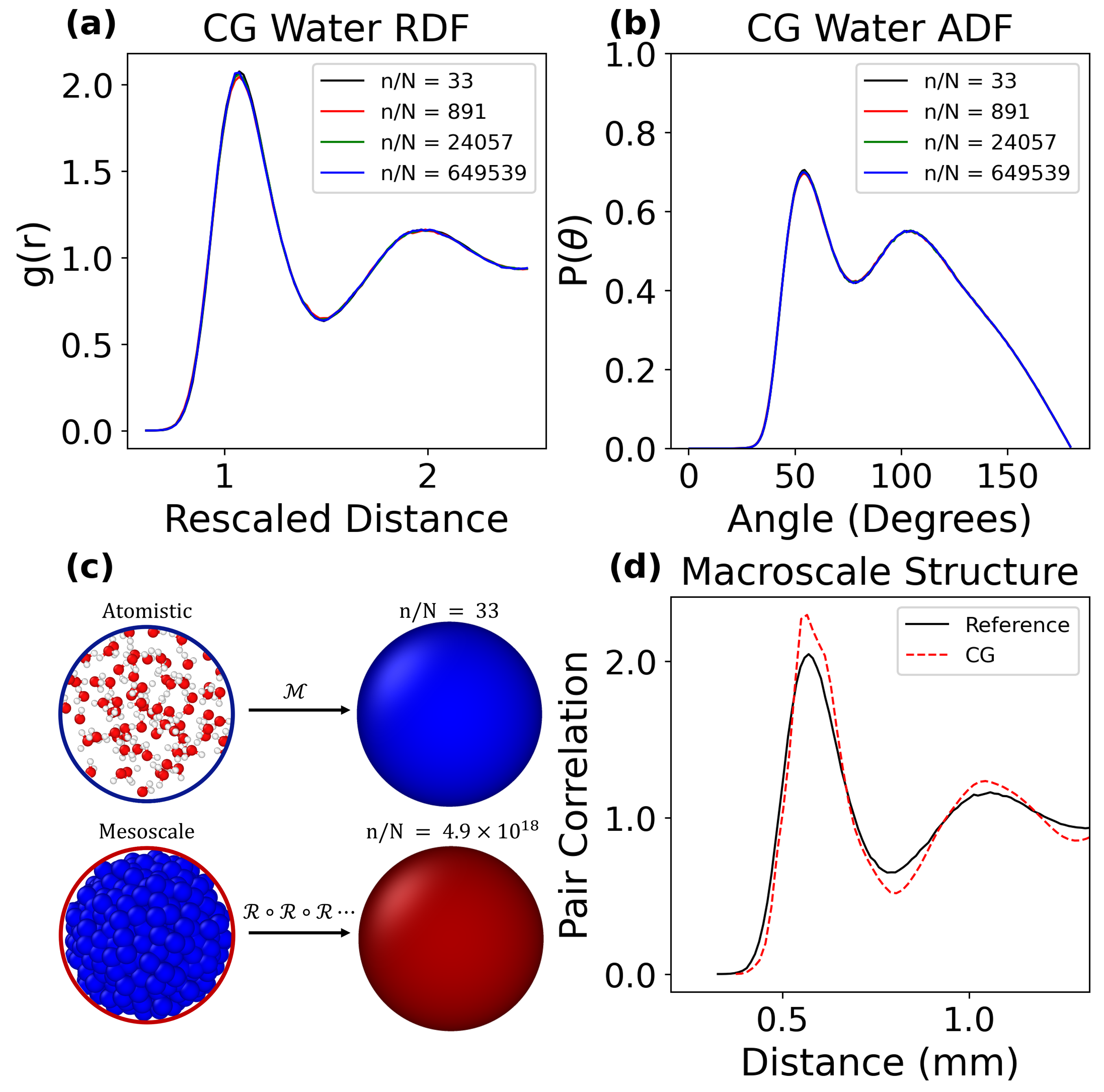}
        \caption{(a) RDFs and (b) ADFs for several repeated CG operations. (c) Iterative CG procedure is shown. Atomistic simulations are conducted to obtain mesoscale representation (upper). Self-similarity enables repeated coarsening operations to reach macroscale degrees of coarsening (lower). (d) Macroscale CG simulation results are shown.}
        \label{fig:renormalization_results}
\end{figure}

Since the number of particles is retained at each iteration, we denote the operator $\mathcal{R}$ to act on the space of CG potentials at a given particle number such that a CG iteration can be summarized as $\mathcal{R}[U_n] = U_{n+1}$.
While the iterative training procedure discussed here provides a systematic route for accessing behavior at any length scale, it is worth considering the properties of $\mathcal{R}$ in discovering potential commonalities between scales. 
Remarkably, we find that structural correlations are preserved at each CG operation, as shown in Figs. \ref{fig:renormalization_results}(a) and \ref{fig:renormalization_results}(b).
In the language of renormalization group theory, the CG potential, approximated by ML potentials, constitutes an infrared fixed point, wherein $\mathcal{R}[U_n] \approx U_n$.

These findings suggest a dramatically simpler approach to hierarchical coarse graining in which a single model can be employed across a breadth of physical scales.
To demonstrate the power of this approach, we consider the $n = 12$th step of this iterative CG operation.
A brute-force approach to reaching the 12th CG iteration would necessitate the development of 11 previous models. 
Instead, utilizing the fixed-point property of the CG potential, we make the approximation $U_{1}(\mathbf{R}/\xi) \approx U_{12}(\mathbf{R}/\xi^{12})$.
Extrinsic properties, in this case the blob mass, are then augmented to appropriately reflect the degree of coarsening at that scale.

Utilizing this rescaling approach, we are able to simulate the $n = 12$th iteration, corresponding to roughly 0.03 mol of water molecules.
Simulating this 0.03 mol equivalent (3456 particles) for 2 seconds of simulation time costs 10.5 hours on a single NVIDIA A100-SXM4 GPU.
Conversely, for the atomistic equivalent, which we largely retain the accuracy of, would constitute $5.4 \times 10^{22}$ atoms and, ignoring clear memory issues, would cost approximately $4.0 \times 10^{22}$ hours on the same GPU to simulate for the same timescales, which is orders of magnitude longer than the age of the universe.
This gives our CG model a $4.0 \times 10^{21}$ order of magnitude computational cost savings.

The results of this simulation, as shown in Fig. \ref{fig:renormalization_results}(d), have the following interpretation.
The Voronoi mapping for the CG water blobs probes the local density of the fluid, averaged across a length scale on the order of the blob radius, and the resulting CG beads are in essence a Lagrangian discretization of the fluid. 
The pair correlation of these blobs then examines the correlation in this spatially averaged density and, emphatically, not the atomistic density.
This interpretation then allows correlations along the millimeter length scale to be physically meaningful.
Consequently, these simulation results demonstrate that the ML CG model developed via ISM retains the structural correlations of heavily coarsened representations of liquid water even at millimeter length scales, extending bottom-up molecular modeling into the macroscopic regime.


\section{Discussion}

A significant limitation of bottom-up CG modeling has been the lack of algorithmic routes to develop high-fidelity models using extremely coarse representations, either through brute-force force matching at extremes or hierarchical CG strategies.
As we have emphasized throughout this work, pushing CG modeling beyond the $\mathcal{O}$(10s) atoms-per-site resolution generally necessitates non-linear dynamical CG mappings to achieve accurate representations.
Applying force-based ML training methods to bridge scales hierarchically poses severe challenges due to the data-inefficiency of the FM method and the computationally prohibitive Jacobian calculations which arise in non-linearly mapped forces.
We have demonstrated here that the ISM method can be used to dramatically reduce the data required to train conventional CG models and can be employed to develop high-fidelity CG models wherein CG sites comprise hundreds to thousands of atoms, made possible through non-linear mapping.
We have further established that models can be trained recursively via ISM to establish a hierarchy of CG models from microscopic physics to macroscale phenomena.


In cases where CG forces are accessible, we find that training to both ISM and FM losses together yields superior models to FM alone. 
Interestingly, ISM and FM loss functions both share the same minimum under complete sampling, yet in practice they emphasize complementary aspects of the model.
Whereas FM directly matches coarse-grained forces and can overfit statistical noise in the force estimates, ISM instead evaluates the learned coarse-grained distribution, naturally discouraging overfitting force noise.
Conversely, the forces of FM provide a physical regularization over the CG model space for ISM's unsupervised learning.
Our observation that a hybrid loss of FM and ISM may produce models more accurate than either method alone has also been observed when employing FM with the closely related denoising score-matching method.\cite{durumeric2026dsm} 
Given the complementary nature of these losses, it is likely that development of state-of-the-art CG models will necessitate hybrid loss functions which target the underlying CG distribution using different approaches.

To truly push the scales of CG modeling, multi-molecular, non-linear CG mappings are a necessity.
We have demonstrated in this work that ISM avoids the complications of employing such mappings for ML training.
Using these ISM models, we are able to capture mesoscale physics wherein a single CG site captures the collective behavior of multiple molecules.
A truly multiscale approach to CG modeling will necessitate further modification to the mappings in this work to incorporate variables beyond spatial coordinates in order to reincorporate fine-grained but pivotal information within the CG model.\cite{jin2026ucg}
For example, CG variables representing the chemical composition of mesoscale CG sites or which account for chemical reactivity within a CG site are a natural next step from this work.
Sub-particle variables that express the changes in these states can be coupled to the inter-particle forces to properly capture long range phenomena such as transport and heterogeneous reaction fronts,
enabling bottom-up modeling of a wide range of macroscopic phenomena.\cite{strachan2005energy,antillon2015mesoscale,antillon2014coarse}

Our findings of a fixed point in recursive ML CG modeling offer a promising approach in bridging atomistic and macroscale, continuum models. 
We have presented, to our knowledge, the first observation of self-similarity in iterative ML CG modeling. 
These findings warrant further research into whether our hierarchical CG approach extends across various phases and compositions of matter.
A particularly promising application of the scale-invariant ML CGs developed in this work is to enable an active learning framework\cite{alf2026grizzi} wherein meso- or macroscale simulations are conducted which may prompt on-the-fly updates to the ML CG directly through ground truth atomistic calculations and re-training strictly at the microscopic scale.


The fundamental challenge in bridging atomistic physics to the mesoscale and macroscale models is the accurate temporal representation of CG dynamics.
That is, the time evolution upon the CG potential alone often artificially accelerates the dynamics, leading to erroneous calculations of transport properties such as diffusion.
Capturing CG dynamics necessitates accurate representation of the missing frictional forces responsible for correcting dynamics,\cite{zwanzig1960}
however, the development of ML methods which capture such forces or accurately rescale dynamics is a growing research direction.\cite{lyu2023cg, wang2025dynamics}

Overall, here, we have developed an agile route for the development of bottom-up models that retain near-atomistic accuracy of critical phenomena while reducing the computational cost by several to tens of orders of magnitude.
The ISM CG model training routine we report here is material/system agnostic and can be applied to effectively any system in which atomistic simulations can resolve the necessary underlying physics.
Our framework is based on a deep, message passing neural network architecture that can be readily augmented to append additional chemical and physical descriptors such as composition, temperature, or orientation.
Hence, our work establishes a viable route toward bottom-up scale bridging across a wide range of scientific disciplines, enabling the development of self-consistent models at multiple resolutions for rigorous multiscale simulations of complex phenomena.

\section{Methods}

\subsection{Atomistic and Coarse-Grained Simulations}

All simulations were conducted with the LAMMPS software package.\cite{thompson2022lammps}
For $\text{C}_{60}$ simulations, we utilize a Lennard-Jones interaction potential and a harmonic bonding potential.\cite{weiss2008carbon}
Simulations consisted of 1372 molecules.

Atomistic simulations of water were conducted at a density of $\rho = 1 \ \text{gm}/\text{cm}^3$ using the SPC/Fw force-field at $T = 300$ K.\cite{wu2006spcfw}
Simulations consisted of 4224 water molecules.

To enable sampling of a large enough bilayer patch such that a single CG sites can feasibly represent an entire patch segment of a lipid bilayer, we utilize the Martini v2.0 CG force-field in generating the reference data for the mesoscale CG model training.\cite{marrink2007martini}
Reference simulations of the Martini CG lipid patch consisted of 4608 lipids at $T = 300$ K.
For all atomistic and CG simulations, reference statistics were calculated under NVT conditions.

\subsection{Machine-Learned Potential Training}

All CG models developed in this work employed the Hierarchically Interacting Particle Neural Network with Tensor Sensitivity (HIP-NN-TS) architecture.\cite{chigaev2023hipnn} 
For force-based training, an equal weighting between root mean squared error and mean absolute error was used. 
For hybrid FM+ISM models, ISM was weighted over the FM model by a factor of $|1/\beta^2|$.
We report a common set of hyperparameters and cutoffs which were used for all ML CG models in Tables S1 and S2.
For all ML CG models developed, a repulsive prior term is added to the potential to discourage sampling in the hard-wall interaction region.

\subsection{Machine-Learned Coarse-Grained Simulations}

CG simulations of $\text{C}_{60}$ were conducted under the NVT ensemble for 600000 timesteps using a timestep of 5 fs and a Langevin damping constant of 0.1 ps. A mass equivalent to the total mass of a $\text{C}_{60}$ molecule was chosen.

Bulk CG simulations of water at the atomistic reference density were conducted under the NVT ensemble for 300000 timesteps using a timestep of 10 fs. 
A damping constant of 1 ps was used.

CG DOPC simulations were conducted under the NVT ensemble using a timestep of 25 fs and a damping constant of 0.1 ps. 

Macroscale water simulations were conducted for 2 seconds using a timestep of 0.5 $\mu$s. A damping constant of 5 $\mu$s was employed in the simulation of 3456 macroscale CG blobs in a cubic box with a length of 8.02 mm. 

\section{Acknowledgments}

Funding for this project was provided by the Advanced Simulation Computing (ASC) Program through the Nicholas C. Metropolis Postdoc Fellowship in Computational and Computer Sciences. Partial funding was provided by the Advanced Simulation and Computing Physics and Engineering Models project (ASC-PEM). This research used resources provided by the Los Alamos National Laboratory (LANL) Institutional Computing Program. This work was supported by the U.S. Department of Energy (DOE) through LANL, which is operated by Triad National Security, LLC, for the National Nuclear Security Administration of the U.S. Department of Energy (Contract No. 89233218CNA000001). Approved for unlimited release: LA-UR-26-28448.

\bibliography{references}

\end{document}


\maketitle
\vspace{-1cm}
\tableofcontents

\section{Implicit Score-Matching Theory}
We assume thermodynamic equilibrium such that the configurational distribution of the atomistic degrees of freedom, $\mathbf{r}$, is, up to a constant, $\log p_{AA}(\mathbf{r}) \propto -\beta u(\mathbf{r}) $ where $\beta = 1/k_BT$.
Bottom-up CG modeling approaches aim to capture the underlying CG potential over the $N < n$ degrees of freedom, $\mathbf{R}$, denoted as $U(\mathbf{R})$.
This potential defines the CG distribution, up to a constant, as $\log p_{CG}(\mathbf{R}) \propto-\beta U(\mathbf{R})$.
To define the forces of the CG potential, we first define the force-mapping operator as
\begin{equation}\label{Eq. 1}
    \boldsymbol{\Xi}(\mathbf{f}(\mathbf{r}))  = \mathbf{D}^{-1}\mathbf{J}\mathbf{f}(\mathbf{r}) + \frac{1}{\beta} \nabla_{\mathbf{r}} \cdot(\mathbf{D}^{-1} \mathbf{J}).
\end{equation}
In Eq. \eqref{Eq. 1}, we have defined the Jacobian of the mapping transformation, $\mathbf{J}_{ij} = \partial\mathbf{R}_i/\partial\mathbf{r}_j$, and the induced metric, $\mathbf{D} = \mathbf{J}\mathbf{J}^T$.
The forces of the CG potential, $\boldsymbol{\mathcal{F}}_I(\mathbf{R}) = -\nabla_I U(\mathbf{R})$, can then be defined as
\begin{equation}\label{Eq. 2}
    \boldsymbol{\mathcal{F}}_I(\mathbf{R}) = \mathbb{E}[\boldsymbol{\Xi}_I (\mathbf{f}(\mathbf{r})) |\boldsymbol{\mathcal{M}}(\mathbf{r}) = \mathbf{R}].
\end{equation}
From the ground truth forces defined in Eq. \eqref{Eq. 2}, one may define a variational residual from which optimal CG potentials can be obtained.
\begin{equation}\label{Eq. 3}
    \mathcal{L}(\theta)  = \frac{1}{3N}\sum_{I=1}^N \mathbb{E}_{\mathbf{R} \sim p} [ \lVert \boldsymbol{\mathcal{F}}_I(\mathbf{R}) + \nabla_I  U_\theta(\mathbf{R})\rVert^2 ].
\end{equation}
Acquiring these forces, however, is complicated by the necessity in evaluating the conditional expectation of Eq. \eqref{Eq. 2} 
Crucially, the mapped force is an unbiased estimator of the ground truth force, 
\begin{equation}\label{Eq. 4}
    \boldsymbol{\Xi}_I (\mathbf{f}(\mathbf{r})) = \boldsymbol{\mathcal{F}}_I(\mathbf{R}) +\boldsymbol{\eta}(\mathbf{r}),
\end{equation}
where $\boldsymbol{\eta}$ is a force-fluctuation term with zero conditional mean, $\mathbb{E}[\boldsymbol{\eta}(\mathbf{r}) |\boldsymbol{\mathcal{M}}(\mathbf{r}) = \mathbf{R}] = 0$.
This identity enables regression onto the mapped atomistic forces in lieu of the ground truth force during training according to the FM method described in the main text.

We next provide a complete derivation of the ISM loss function.
From the divergence theorem, for any differentiable vector field of the CG configuration space $\mathbf{A}(\mathbf{R})$ the following holds
\begin{equation}\label{Eq. 5}
    \mathbb{E}_{\mathbf{R} \sim p} [ \nabla \cdot (p(\mathbf{R}) \cdot \mathbf{A}(\mathbf{R}))] = 0 .
\end{equation}
We next briefly derive the Stein identity from Eq. \eqref{Eq. 5} by the product rule and the log-derivative trick,
\begin{equation}\label{Eq. 6}
    \mathbb{E}_{\mathbf{R} \sim p_{CG}} [ \nabla \cdot \mathbf{A}(\mathbf{R})]  = - \mathbb{E}_{\mathbf{R} \sim p_{CG}} [ \mathbf{A}(\mathbf{R}) \cdot \nabla \log p_{CG}(\mathbf{R})] 
\end{equation} 
Then, specifically for the Boltzmann distribution of the CG PMF, we have from Eq. \eqref{Eq. 6}
\begin{equation}\label{Eq. 7}
    \mathbb{E}_{\mathbf{R} \sim p_{CG}} [ \nabla \cdot \mathbf{A}(\mathbf{R})]=  \beta \cdot\mathbb{E}_{\mathbf{R} \sim p_{CG}} [  \mathbf{A}(\mathbf{R}) \ \cdot \nabla U(\mathbf{R})].
\end{equation}
We now consider the CG model potential, $U_\theta(\mathbf{R})$, and let $\mathbf{A}(\mathbf{R}) = \nabla U_\theta(\mathbf{R})$. Then, by the Stein identity of Eq. \eqref{Eq. 3},
\begin{equation}\label{Eq. 8}
    \mathbb{E}_{\mathbf{R} \sim p_{CG}} [ \nabla^2 U_\theta (\mathbf{R})  ] = \beta \cdot \mathbb{E}_{\mathbf{R} \sim p_{CG}} [ \nabla U_\theta (\mathbf{R}) \cdot  \nabla  U(\mathbf{R})].
\end{equation}
Next, we consider the loss function of Eq. \eqref{Eq. 3}, and its form as a quadratic into three terms,
\begin{equation}\label{Eq. 9}
\begin{aligned}
    \mathcal{L}(\theta)  =& \frac{1}{3N}\Big [\sum_{I=1}^N\mathbb{E}_{\mathbf{R} \sim p_{CG}} [ \lVert U_\theta(\mathbf{R})\rVert^2 ]\\
    &- 2 \cdot \sum_{I=1}^N \mathbb{E}_{\mathbf{R} \sim p_{CG}} [ \nabla U_\theta (\mathbf{R}) \cdot  \nabla  U(\mathbf{R}) ] + \sum_{I=1}^N \mathbb{E}_{\mathbf{R} \sim p_{CG}} [ \lVert U(\mathbf{R})\rVert^2 ]\Big].
\end{aligned}
\end{equation}
The latter term is a constant we will drop, renaming the loss function $\mathcal{L}_{ISM}$ at this point. The middle term can be rewritten according to Eq. \eqref{Eq. 8}. 
\begin{equation}\label{Eq. 10}
    \mathcal{L}_{ISM}(\theta)  = \frac{1}{3N}\Big [\sum_{I=1}^N \mathbb{E}_{\mathbf{R} \sim p_{CG}} [ \lVert \nabla U_\theta(\mathbf{R})\rVert^2 ]- \frac{2}{\beta} \cdot  \sum_{I=1}^N\mathbb{E}_{\mathbf{R} \sim p_{CG}} [ \nabla^2 U_\theta (\mathbf{R}) ] \Big ].
\end{equation}
Lastly, we employ the identity $p(\mathbf{R}) = \mathbb{E}_{\mathbf{r}\sim p_{AA}} [\delta (\boldsymbol{\mathcal{M}}(\mathbf{r})-\mathbf{R})]$ such that the expectations may be evaluated over the atomistic configurations,
\begin{equation}\label{Eq. 11}
    \mathcal{L}_{ISM}(\theta)  = \frac{1}{3N} \Big [\sum_{I=1}^N\mathbb{E}_{\mathbf{r} \sim p_{AA}} [ \lVert \nabla U_\theta(\boldsymbol{\mathcal{M}}(\mathbf{r}))\rVert^2 ]- \frac{2}{\beta} \cdot \sum_{I=1}^N \mathbb{E}_{\mathbf{r} \sim p_{AA}} [ \nabla^2 U_\theta (\boldsymbol{\mathcal{M}}(\mathbf{r})) ] \Big ].
\end{equation}
which is precisely the implicit score-matching loss function.

It is worth noting that the constant by which $\mathcal{L}$ and $ \mathcal{L}_{ISM}$ differ is the average of the square of the PMF force, which is a non-negative value. 
Since the optimal minimum of $\mathcal{L}$ is zero, this then implies that the optimal value of $\mathcal{L}_{ISM}$ is a negative number.

In practice, the direct evaluation of the CG model Hessian is a significant computational bottleneck.
To circumvent this issue, we employ the Hutchinson trace estimator, wherein for a matrix, $\mathbf{B}$,
\begin{equation}\label{Eq. 12}
    \operatorname{Tr} \mathbf{B} = \mathbb{E}_{\mathbf{v} \sim p(\mathbf{v})}[\mathbf{v}^T\mathbf{B}\mathbf{v}],
\end{equation}
where the vectors $\mathbf{v}$ are distributed such that
\begin{equation}\label{Eq. 13}
    \mathbb{E}_{\mathbf{v} \sim p(\mathbf{v})}[\mathbf{v}\mathbf{v}^T] = \mathbf{I}.
\end{equation}
This reduces the calculation of the trace of the Hessian matrix in Eq. \eqref{Eq. 11} to evaluation of numerically cheaper Hessian vector products.
Practically, then, for a batch size $M$ and $K$ repeated random vector probes we implement the following form of the ISM loss function,
\begin{equation}\label{Eq. 14}
    \mathcal{L}_{ISM}(\theta;M,K)  = \frac{1}{3N}\Big[\frac{1}{M}\sum_{I=1}^N \sum_{b=1}^M  \lVert \nabla U_\theta(\boldsymbol{\mathcal{M}}(\mathbf{r}_b))\rVert^2 - \frac{2}{\beta} \cdot \frac{1}{M}\cdot  \sum_{b=1}^M \frac{1}{K}\cdot \sum_{k=1}^K \mathbf{v}^T_k\nabla^2 U_\theta (\boldsymbol{\mathcal{M}}(\mathbf{r}_b))\mathbf{v}_k  \Big ].
\end{equation}
Throughout this work, we employ $K = 1$ vector probes during training.

\section{Additional Information on Atomistic and Coarse-Grained Simulations}

Atomistic simulations of $\text{C}_{60}$ were conducted by first minimizing the energy of the system for 10000 iterations. 
Next, the system was briefly ran under the NPT ensemble for 50 ps with a timestep of 1 fs at 1 atm.
A barostat coupling constant of 1 ps and a Langevin damping constant of 0.1 ps were employed to maintain temperature and pressure, respectively.
The cell was then replicated to produce a system of 1372 $\text{C}_{60}$ molecules.
The system was then ran again under the NPT ensemble for 100 ps, followed by 10 ns of NVT simulation, from which the last 8 ns was utilized as the training dataset.

A timestep of 0.2 fs was employed for NVT simulations of 4224 water molecules for roughly 22.5 nanoseconds using a Langevin damping constant of 0.1 ps.
Simulations of a liquid-vapor interface were also performed to comprise the training dataset for droplet simulations.
This system was initialized by taking the water system at a cell size of $2.5 \times 2.5 \times 2.5$ $ \text{nm}^3$ to a cell size of $2.5 \times 2.5 \times 20$ $\text{nm}^3$. 
The liquid-vapor interface was then simulated under NVT conditions for 1.5 ns.

Reference simulations of the Martini CG lipid patch were obtained in the following manner. 
First a small lipid patch of 288 lipids was energy minimized for a maximum number of 100000 iterations.  
The patch was then simulated under NVT conditions at $T = 600$ K for roughly 20000 timesteps at a timestep of 30 fs with a damping constant of 1 ps.
This was then followed by simulation under NPT conditions for 150000 timesteps in which the temperature was cooled to 300 K and pressure was maintained from 200 to 1 atmosphere. A barostat coupling constant of 1 ps was used. 
An equilibration run at $T = 300$ K and $P = 1$ was then conducted for 100000 timesteps.
After equilibration, the membrane patch was then replicated to constitute 4608 lipids, and a simulation under NVT conditions was ran for roughly 20 million timesteps with a timestep of 30 fs.

\section{Additional Information on Coarse-Grained Models}
Hyperparameter values shared among all ML models developed in this work are shown in Table. \ref{tab:hyperparameters}.
The Adam optimizer was implemented during training. 
The batch size was doubled every 10 consecutive epochs without improvement in the validation loss and training was terminated after 20 consecutive epochs.
Training proceeded in total for no more than 200 epochs.
For all CG models in this work, a potential of the form $U_\theta = U_{prior} + U_{HIP-NN-TS}$ was used. We define $U_{prior}$ as
\begin{equation}\label{Eq. 15}
    U_{prior} = se^{-t\cdot (r - r_c)}.
\end{equation}
For the $\text{C}_{60}$ and DOPC models, a values of $s = 2.0$ kcal/mol $\mathring{\text{A}}$ was used. 
For these two models, $r_c$ was chosen from the RDF statistics to be the first $r$ value where $g(r) > 0.0005$.
$s$ was then chosen from the average force value. 
For the iterative CG models developed, $r_c$ was chosen from the RDF statistics to be the first $r$ value where $g(r) > 0.005$. The value of $s$ was then chosen to be the the Boltzmann-inverted value of the first nonzero RDF bin. 
For the iterative CG models, the value of $t$ was set to 4.0 and for all other CG models the value of $t$ was set to 1.0.
Interaction cutoff values for all models are shown in Table. \ref{tab:cutoffs}. 

RDF error was quantified according to the following metric, $\Delta g(r) = \frac{1}{r_{max}}\int^{r_{max}} \text{d}r |g(r) - g_\text{ref}(r)|$, where $r_{max}$ is a system specific cutoff.
ADF error was similarly quantified according to the following relation, $\Delta P(\theta) =  \frac{1}{\theta_{max}} \int^{\theta_{max}} \text{d}\theta |P(\theta) - P_\text{ref}(\theta)|$, where ${\theta_{max}} = 180^\circ$ and angles are obtained for all neighboring triplets within the first solvation shell.
\begin{table}[!htbp]
  \caption{Hyperparameters for ML CG FFs.}
  \label{tab:hyperparameters}
    \begin{tabular}{ll}
        Hyperparameters  & Value \\
        $\verb|n_features|$  & 128\\
        $\verb|n_sensitivities|$ & 20 \\
        $\verb|n_interaction_layers|$ & 1\\
        $\verb|n_atom_layers|$ & 3\\
        $\verb|sensitivity_type|$ & inverse \\
        $\verb|resnet|$ & true \\
        $\verb|batch_size|$ & 1 \\
        $\verb|learning_rate|$ & 1e-3
        \end{tabular}
\end{table}

\begin{table}[!htbp]
  \caption{Interaction cutoffs for ML CG FFs.}
  \label{tab:cutoffs}
    \begin{tabular}{llll}
        Model  & $\verb|dist_min| \ (\mathring{\text{A}})$  & $\verb|dist_soft_min| \ (\mathring{\text{A}})$ & $\verb|dist_soft_max| \ (\mathring{\text{A}})$   \\
        $\text{C}_{60}$   & 8.05 & 19.0 & 20.0 \\
        Water (n/N = 12.4) & 4.0625 & 23.0 & 24.0  \\
        Water (n/N = 16.5) & 4.6875 & 23.0 & 24.0  \\
        Water (n/N = 33.0) & 7.1875 & 23.0 & 24.0  \\
        Water (n/N = 40.2) & 7.6042 & 23.0 & 24.0  \\
        DOPC & 37.5 & 99.0 & 100.0\\
        \end{tabular}
\end{table}

\subsection{Buckminsterfullerene}

\begin{figure}[H]
        \centering
        \includegraphics[width=1\linewidth]{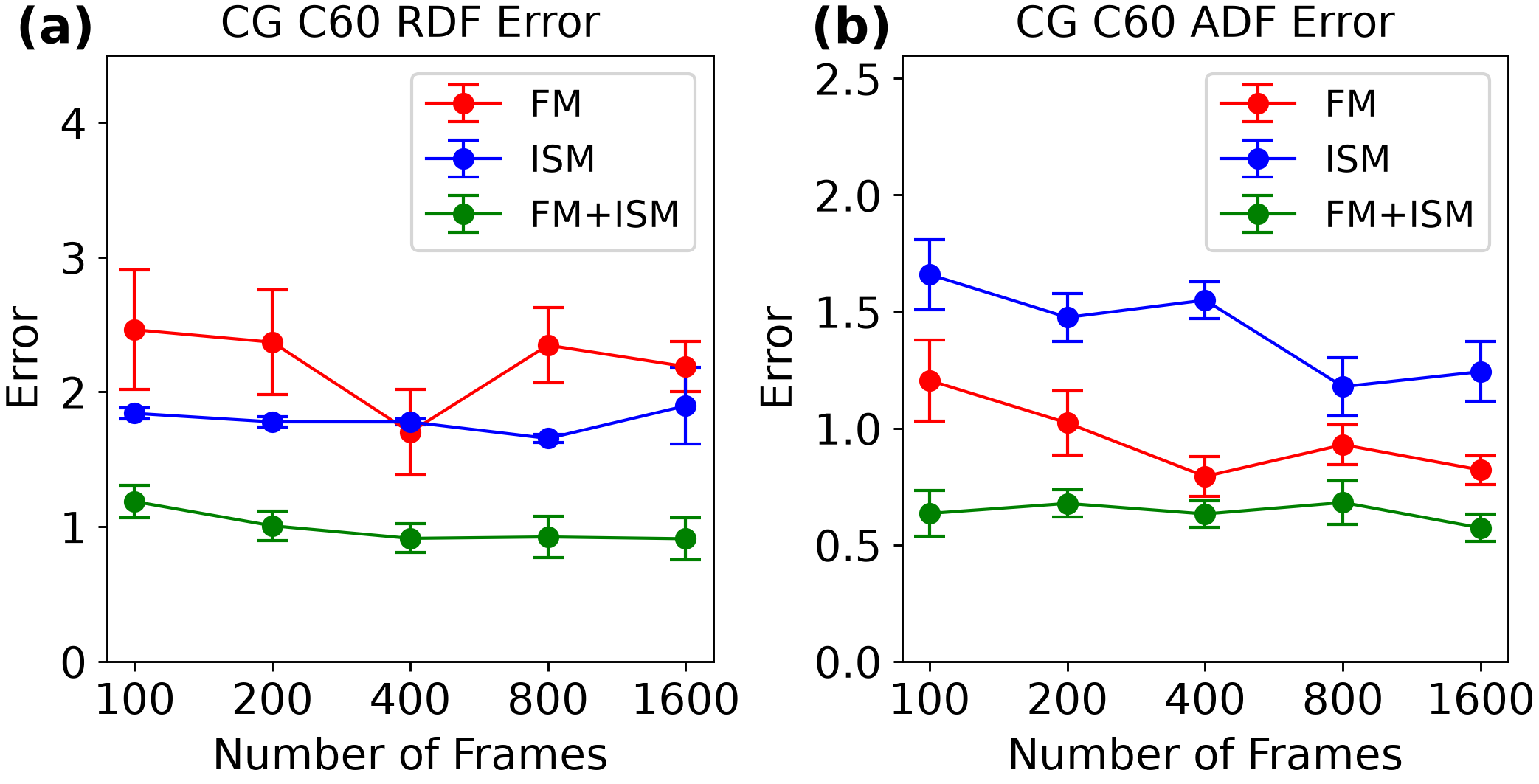}
        \caption{(a) RDF and (b) ADF errors for all CG C60 models are shown.}
        \label{fig:C60_errors}
\end{figure}

Model errors across a range of dataset sizes are shown in Fig. \ref{fig:C60_errors}.
Loss metric for all developed model type s are shown in Table. \ref{tab:c60loss}.

\begin{table}[h]
    \caption{Loss metrics for $\text{C}_{60}$ ML CG models trained.}
    \label{tab:c60loss}
    \begin{tabular}{c c c}
\hline
Model & FM Loss (kcal/mol $\mathring{A}$) & ISM Loss (kcal/mol $\mathring{A}$) \\
\hline
FM & $ 15.62 \pm 0.01$ & $-69.81 \pm 0.16$ \\
ISM & $ 15.86 \pm 0.02$ & $-70.46 \pm 0.13$ \\
FM+ISM & $ 15.71 \pm 0.02$ & $-70.57 \pm 0.14$ \\
\hline
\end{tabular}
\end{table}

\subsection{Mesoscale Coarse-Graining of Water}
\subsubsection{Mapping}
The soft-Voronoi mapping employed is mathematically defined from the fixed-point relation, 
\begin{equation}\label{Eq. 16}
    \mathbf{R} = \boldsymbol{\mathcal{M}}_1(\mathbf{R}, \boldsymbol{{\mathcal{M}}}_0(\mathbf{r}))
\end{equation}
where $\boldsymbol{{\mathcal{M}}}_0$ is the center-of-mass mapping operator for water molecules. The blob mapping is defined as
\begin{equation}\label{Eq. 17}
    \boldsymbol{\mathcal{M}}_{1,I}(\mathbf{R},\boldsymbol{{\mathcal{M}}}_0(\mathbf{r})) = \frac{\sum_i w_{Ii}(|\mathbf{R}_I-\boldsymbol{{\mathcal{M}}}_{0,i}(\mathbf{r})|)\boldsymbol{{\mathcal{M}}}_{0,i}(\mathbf{r})}{\sum_i w_{Ii}(|\mathbf{R}_I-\boldsymbol{{\mathcal{M}}}_{0,i}(\mathbf{r})|)},
\end{equation}
where the particle-to-blob weights are defined as
\begin{equation}\label{Eq. 18}
    w_{Ii}(|\mathbf{R}_I-\boldsymbol{{\mathcal{M}}}_{0,i}(\mathbf{r})|) = \frac{f_{Ii}(|\mathbf{R}_I-\boldsymbol{{\mathcal{M}}}_{0,i}(\mathbf{r})|)}{\sum_If_{Ii}(|\mathbf{R}_I-\boldsymbol{{\mathcal{M}}}_{0,i}(\mathbf{r})|)},
\end{equation}
and we have defined a configurational weighting function, $f$, as
\begin{equation}\label{Eq. 19}
    f_{Ii}(|\mathbf{R}_I-\boldsymbol{{\mathcal{M}}}_{0,i}(\mathbf{r})|) = \frac{1}{2}\cdot \Big ( 1 - \tanh(|\mathbf{R}_I-\boldsymbol{{\mathcal{M}}}_{0,i}(\mathbf{r})| - \ell_N) \Big )\cdot \text{erfc}(|\mathbf{R}_I-\boldsymbol{{\mathcal{M}}}_{0,i}(\mathbf{r})|).
\end{equation}
We have also defined a characteristic length scale we denote as $\ell_N = L/\sqrt[3]{4\pi N/3}$ where $L$ is the side length of the cubic box.
It is clear from Eq. \eqref{Eq. 16} that $\mathbf{R}$ is defined as a function of itself, and must therefore be solved self-consistently.
We employ an iterative update scheme 
\begin{equation}\label{Eq. 20}
    \mathbf{R}^{(k+1)} = \mathbf{R}^{(k)}+ \alpha \cdot (\boldsymbol{\mathcal{M}}(\mathbf{R}^{(k)},\boldsymbol{{\mathcal{M}}}_0(\mathbf{r})) - \mathbf{R}^{(k)})
\end{equation} to obtain a converged set of $\mathbf{R}$.
A random subset of $\boldsymbol{{\mathcal{M}}}_0(\mathbf{r})$ are chosen to act as the initial configurations for $\mathbf{R}$ within the self-consistent procedure.
We set $\alpha = 0.5$ and update to a self-consistency of $|\mathbf{R}^{(k+1)} - \mathbf{R}^{(k)}|\leq 1\text{e}{-5} \ \mathring{\text{A}}$ for up to 5000 iterations, although we do not observe any $\mathbf{R}$ which do not converge within this iteration number. 

We briefly remark on how approximate forces were obtained for FM training of this system.
The instantaneous force on the CG `blobs' can be related to the momentum time derivative $\mathbf{F}_I = d\mathbf{P}_I/dt$.
The CG blob momentum can be expressed, approximately, as 
\begin{equation}\label{Eq. 21}
    \mathbf{P}_I(t) \approx  \frac{\sum_i w_{Ii}(|\mathbf{R}_I-\boldsymbol{{\mathcal{M}}}_{0,i}(\mathbf{r})|)m_i\mathbf{v}_i(t)}{\sum_i w_{Ii}(|\mathbf{R}_I-\boldsymbol{{\mathcal{M}}}_{0,i}(\mathbf{r})|)},
\end{equation}
where $m_i$ and $\mathbf{v}_i$ are the mass of a water molecule and the center-of-mass velocity, respectively.
We note that this ignores contributions to the CG force attributed to the dynamics of the CG mapping operator.
From this expression, we employ a central finite difference to obtain the CG instantaneous force,
\begin{equation}\label{Eq. 22}
    \mathbf{F}_I(t) \approx \frac{\mathbf{P}_I(t+\Delta t) - \mathbf{P}_I(t-\Delta t)}{2\cdot \Delta t}.
\end{equation}
In this work, a $\Delta t$ of 0.2 fs was used.

\begin{figure}[H]
        \centering
        \includegraphics[width=0.8\linewidth]{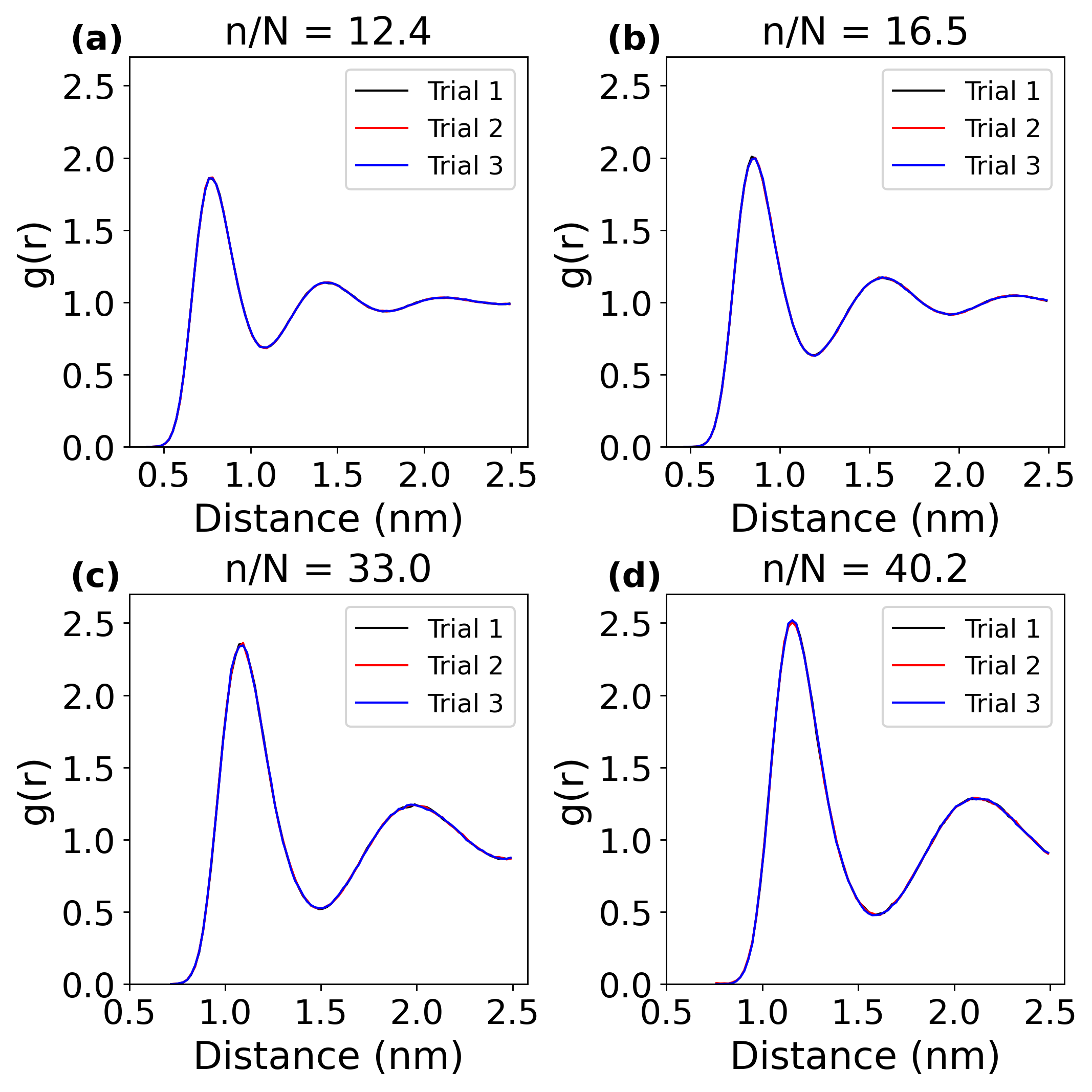}
        \caption{RDFs employing different initial conditions are shown for all CG resolutions.}
        \label{fig:consistency}
\end{figure}

\begin{figure}[H]
        \centering
        \includegraphics[width=0.8\linewidth]{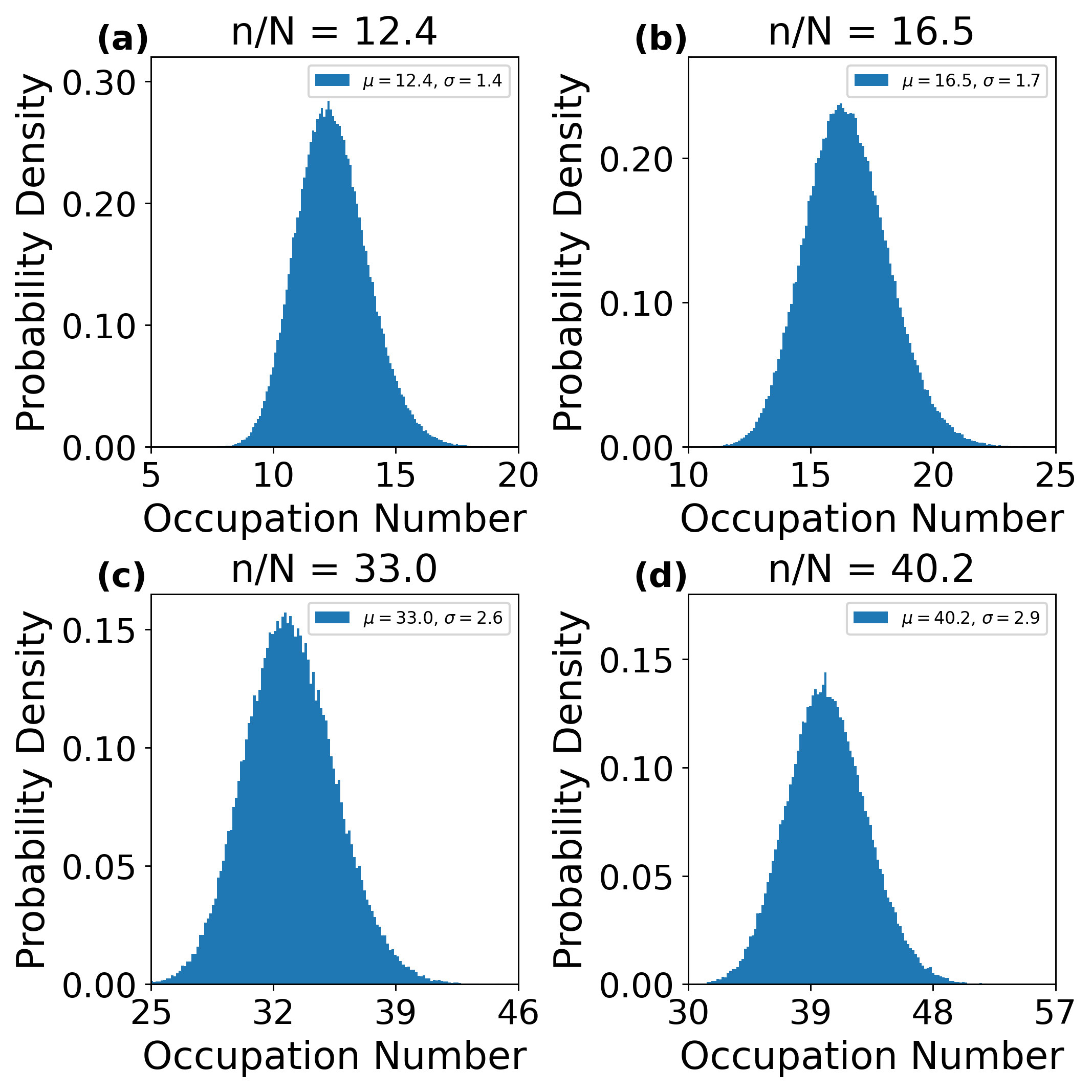}
        \caption{Water molecule occupation number distributions for all  considered CG mappings are shown.}
        \label{fig:occupation}
\end{figure}

We consider the results of mapping from different initial random subsets of $\boldsymbol{{\mathcal{M}}}_0(\mathbf{r})$ to ensure that initial conditions do not significantly affect the solution of this self-consistent mapping problem.
In Fig. \ref{fig:consistency} we show the results of varying the initial conditions for the CG resolutions considered, from which it can be concluded that the mapping procedure of Eqs. \eqref{Eq. 16}--\eqref{Eq. 20} is robust against changes to initial conditions.
Additionally, we report in Fig. \ref{fig:occupation} the variation in the number of water molecules which comprise a CG blob.
We find that the distribution in occupation number is roughly a normal distribution centered around the ratio of water molecules to blobs, indicating that the mapping retains spatial homogeneity in the blob density.

\newpage

\subsubsection{Additional Results}
\begin{figure}[h]
        \centering
        \includegraphics[width=0.8\linewidth]{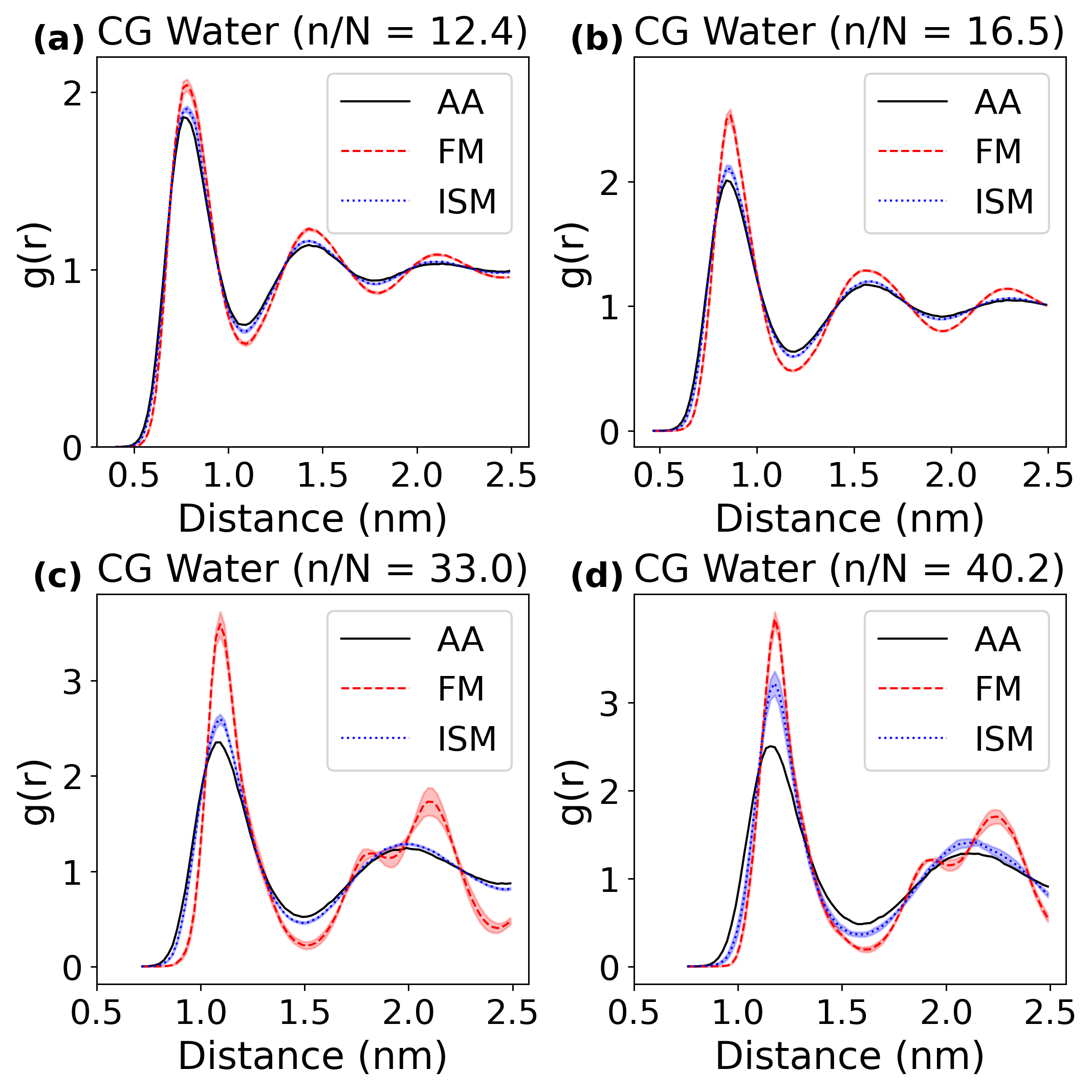}
        \caption{RDFs for all models developed at the considered CG mappings are shown.}
        \label{fig:water_rdf_results}
\end{figure}

\begin{figure}[htpb]
        \centering
        \includegraphics[width=0.8\linewidth]{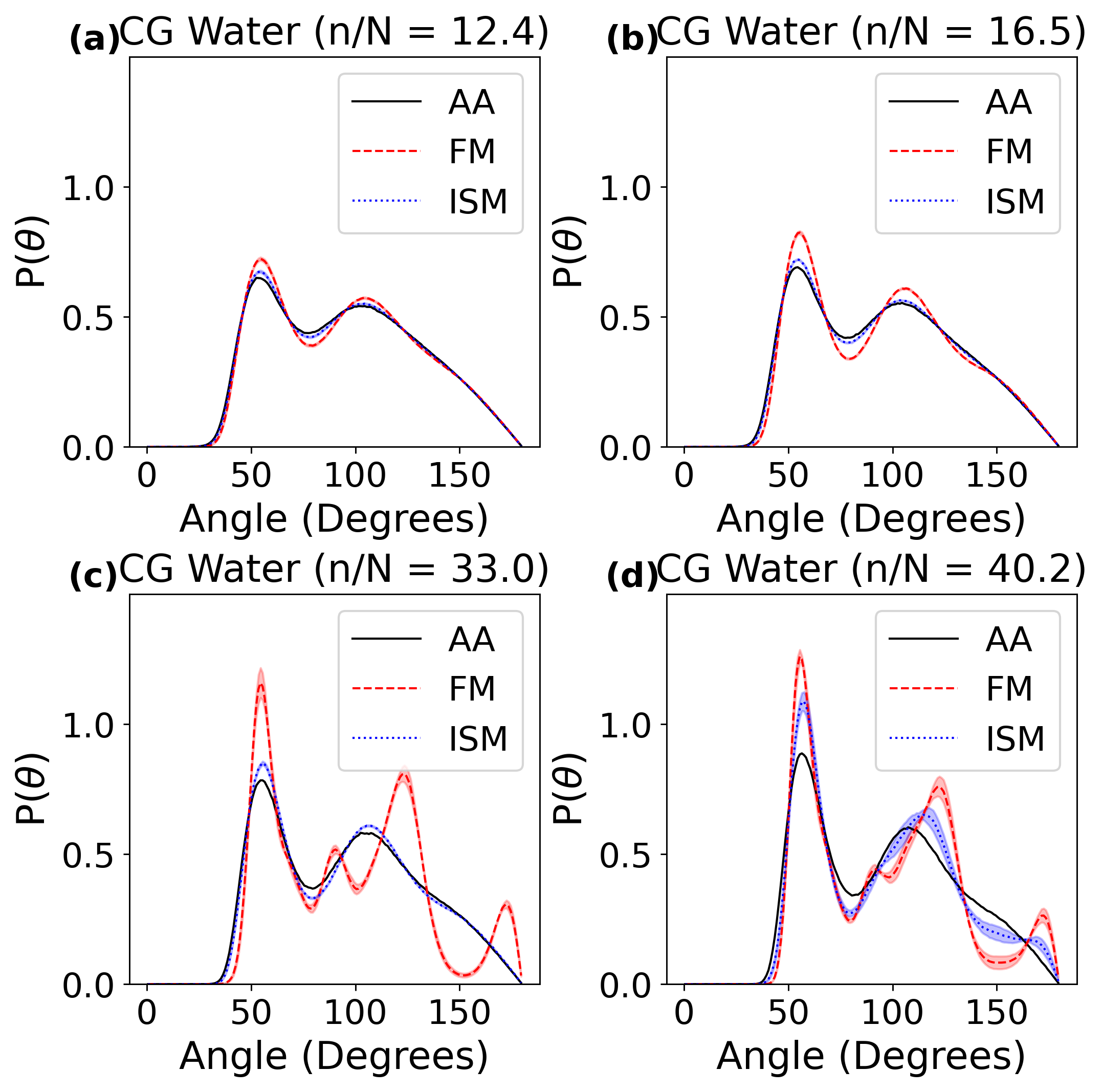}
        \caption{ADFs for all models developed at the considered CG mappings are shown.}
        \label{fig:water_adf_results}
\end{figure}

\begin{figure}[htpb]
        \centering
        \includegraphics[width=0.8\linewidth]{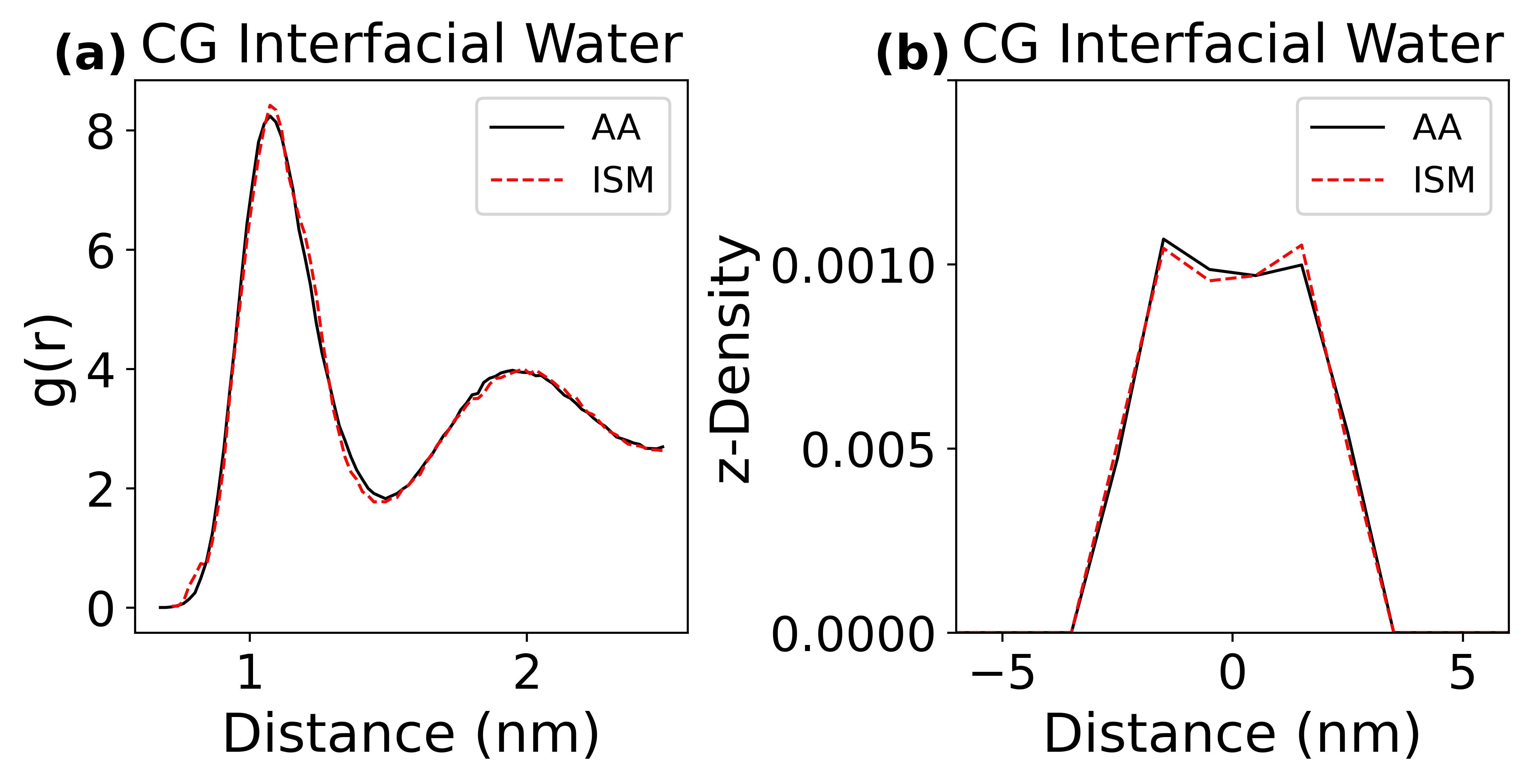}
        \caption{(a) RDF and (b) interfacial density profile for mapped AA and ISM model are shown.}
        \label{fig:water_interface}
\end{figure}

\begin{figure}[htpb]
        \centering
        \includegraphics[width=0.8\linewidth]{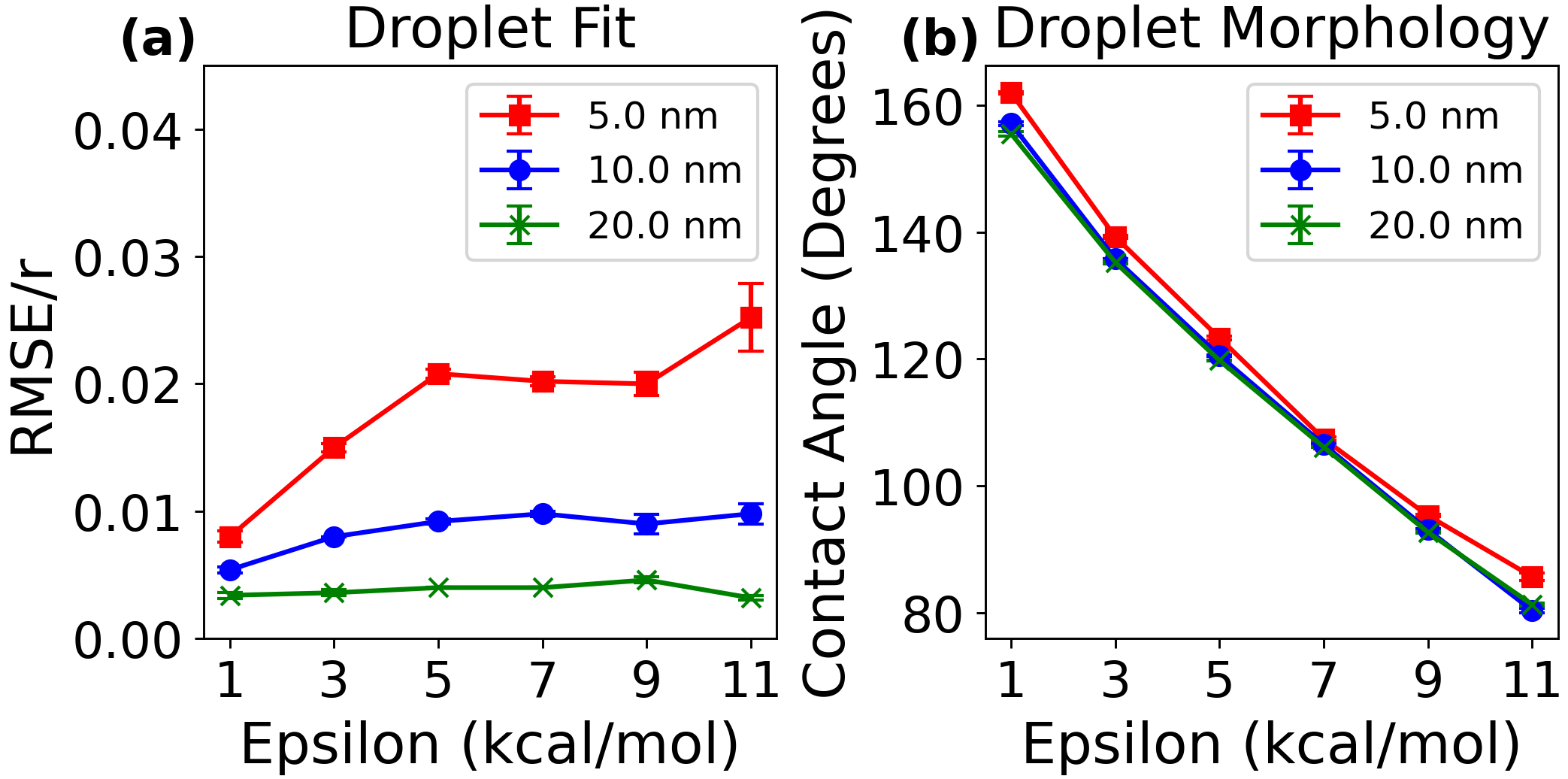}
        \caption{(a) Fits of droplets to the equation of a sphere relative to sphere radius are shown. (b) Variation of contact angle with respect to wall interaction strength are shown.}
        \label{fig:water_droplet}
\end{figure}

RDF and ADF error for all bulk CG water model resolutions are shown in Figs. \ref{fig:water_rdf_results} and \ref{fig:water_adf_results}, respectively. 
The RDF and interfacial density for both mapped AA and ISM models are shown in Fig. \ref{fig:water_interface}.
From the ISM model trained on interfacial water, droplet results are shown in Fig. \ref{fig:water_droplet}.

Kirkwood-Buff solution theory provides a connection between microscopic integrals and bulk properties.
In principle, these calculations provide the same material properties regardless of the simulation resolution.
To this end, we consider as an example the compressibility equation for the isothermal compressibility, $\kappa_T$, of the bulk water system, defined by
\begin{equation}\label{Eq. 23}
   \kappa_T =\lim_{k\to 0} \frac{S(k)}{\rho k_BT},
\end{equation}
where $\rho = n/V$ is the molecular density and $S(k)$ is the structure factor,
\begin{equation}\label{Eq. 24}
   S(k) = \frac{1}{n} \Big \langle \sum_{I=1}^N |O_Ie^{-i \mathbf{k} \cdot \mathbf{R}_I}|^2\Big \rangle_{|\mathbf{k}| = k}.
\end{equation}
We have additionally defined the occupation number for CG site $I$ as $O_I$. 
The occupation number for, e.g., a center-of-mass mapping is trivially defined as $O_I = 1 \ \forall I \in \mathbf{O}$.
For the extremely coarse resolution of the soft-Voronoi mappping considered in this section, the occupation numbers for all CG sites exhibit fluctuations, as shown in Fig. \ref{fig:occupation}. 
The resultant CG model and simulation, however, has removed the occupation number degrees of freedom, and thus do not have the full amount of information required to extract the structure factor and hence predict the isothermal compressibility. 
To remedy this, we introduce a backmapping operation to reconstruct the missing occupation numbers in the form of a generative ML model, 
\begin{equation}
    p(\mathbf{O}|\mathbf{R}) \approx \mathcal{N}(\boldsymbol{\mu}_\theta (\mathbf{R}), \boldsymbol{\sigma}^2_\phi (\mathbf{R})).
\end{equation}

Here, we have introduced two message-passing graph neural networks, $\mu_\theta$ and $\sigma_\theta$ which determine the occupation mean and variance at a given blob configuration. 
The occupation numbers at a given configuration can then be sampled via $\tilde{\mathbf{O}} = \boldsymbol{\mu}_\theta  + \boldsymbol{\sigma}_\phi \circ \boldsymbol{\epsilon}$ where $\epsilon \sim \mathcal{N}(0,1)$. 
The generative backmapping model was trained according to the negative log-likelihood. 
Briefly, the hyperparameters of the model consisted of two message passing operations with a cutoff of 10 $\mathring{\text{A}}$, a 64-dimensional latent dimension, and 16 uniformly spaced Gaussian radial basis functions.
Messages consisted of three linear update blocks and two SiLu operations with a residual update.

\begin{figure}[htpb]
        \centering
        \includegraphics[width=0.8\linewidth]{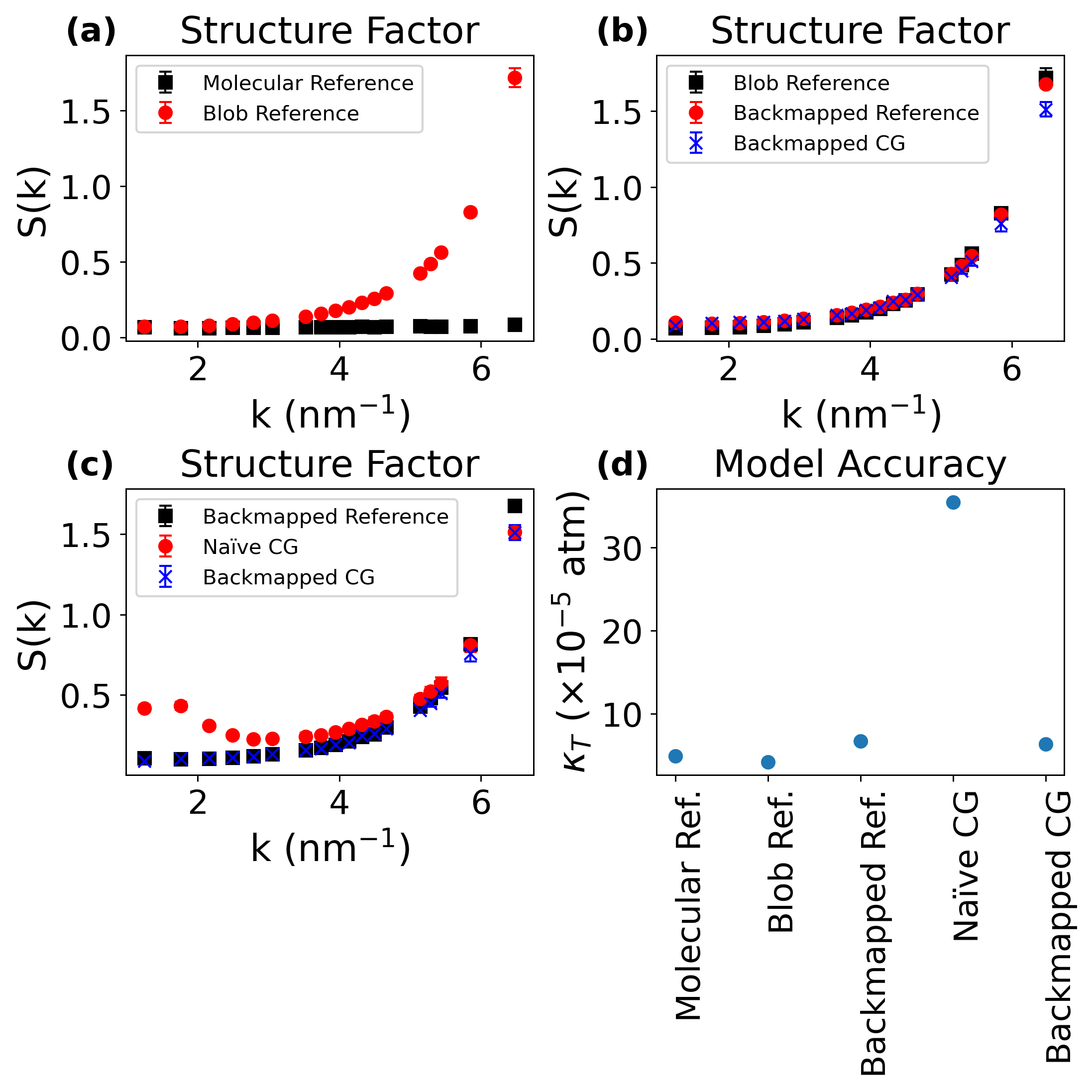}
        \caption{(a) Structure factor of atomistic model mapped at varying resolutions. (b) Structure factors of atomistic model mapped to blob resolution, atomistic model with occupation numbers obtained through the backmapping procedure, and CG model with occupation numbers obtained through backmapping. (c) Structure factors obtained from atomistic and ISM CG models with occupation numbers obtained through backmapping as well as the naïvely obtained ISM CG structure factor. (d) Isothermal compressibilities obtained for all models.}
        \label{fig:water_compressibility}
\end{figure}

We next consider comparisons between the mapped atomistic models and ISM CG model for the $n/N = 12.4$ resolution.
We provide comparisons of the asymptotic behavior of the structure factor for the center-of-mass coarsening and the soft-Voronoi blob mapping when applied to the reference atomistic model in Fig.  \ref{fig:water_compressibility}(a).
We denote these structure factors as Molecular Reference and Blob Reference, respectively.
We emphasize that it is only in the asymptotic limit that the two structure factors should agree.
We next compare the structure factor obtained from the reference blob configurations and occupation numbers extracted from the soft Voronoi mapping against those obtained from the generative backmapping procedure applied to both the reference and CG model blob configurations, which we denote as Backmapped Reference and Backmapped CG, as shown in Fig. \ref{fig:water_compressibility}(b).
Succinctly, Blob Reference obtains the occupation numbers from the bottom-up via $\{\mathbf{R}, \mathbf{O}\} = \mathcal{M}_1 \circ \mathcal{M}_0(\mathbf{r})$ where as the backmapping procedure obtains the occupation numbers from $p(\mathbf{O}, \mathbf{R}) = p(\mathbf{O|R)} p(\mathbf{R})$.
We find generally close agreement amongst all three structure factors, however each leads to slightly different asymptotic behavior and hence variation in the reported isothermal compressibility.
We note that the structure factors obtained by the CG model relies on the quality of both the CG model as well as the backmapping operation, thus the optimal structure factor and isothermal compressibility for the CG model, and hence point of comparison, should be the Backmapped Reference structure factor.
We additionally compare in Fig. \ref{fig:water_compressibility}(c) the structure factor obtained from naïvely assuming a constant blob occupation number, $O_I = n/N$, which we refer to as Naïve CG, against that obtained from the generative backmapping procedure applied to the reference and CG model configurations.
We find strong disagreement in the asymptotic behavior of the naïvely obtained structure factor with all other results, leading to an order of magnitude disagreement in the reported isothermal compressibility, as shown in Fig. \ref{fig:water_compressibility}(d).
This demonstrates the importance of retaining variables which describe the internal behavior or state of the CG sites under aggressive coarsening operations.
We report the values of the isothermal compressibility for these models in Table \ref{tab:compressibility}. 

\FloatBarrier

\begin{table}[htpb]
    \caption{Isothermal compressibilities obtained for each model.}
    \label{tab:compressibility}
    \begin{tabular}{c c }
\hline
Model & $\kappa_T \ ( \times 10^{-5} \text{ atm})$ \\
\hline
Molecular Reference &  4.9 \\
Blob Reference & 4.2 \\
Backmapped Reference & 6.7  \\
Naïve CG & 35.4  \\
Backmapped CG & 6.4  \\

\hline
\end{tabular}
\end{table}

\subsection{Mesoscale Coarse-Graining of a Lipid Bilayer}
\subsubsection{Mapping}
For a bilayer of length $L$ we first define a set of $D \times  D$ fixed mesh grid points,
\begin{equation}\label{Eq. 25}
    \mathbf{\tilde{R}}_{nm} = ((n+0.5)\cdot( L/n), (m+0.5)\cdot( L/m), 0),
\end{equation}
where $n, m \in (0, 1, \cdots, D-1)$. 
The Martini NC3 beads, whose coordinates encompass $\mathbf{r}$ for this CG operation, are used to probe the lateral lipid density around each mesh point, which then constitutes the definition of the CG sites.
We define the mapping operator as
\begin{equation}\label{Eq. 26}
    \boldsymbol{\mathcal{M}}_{nm}(\mathbf{r}) = \mathbf{\tilde{R}}_{nm} + \boldsymbol{\Delta}_{nm}(\mathbf{r}),
\end{equation}
where the influence of lipids local to the defined mesh point are defined as
\begin{equation} \label{Eq. 27}
    \boldsymbol{\Delta}_{nm}(\mathbf{r}) = \frac{\sum_i w_{nm}(\mathbf{r}_i) \mathbf{r}_i}{\sum_i w_{nm}(\mathbf{r}_i) },
\end{equation}
and we have introduced a weighting proximity function
\begin{equation} \label{Eq. 28}
    w_{nm}(\mathbf{r}_i) = e^{-((x_i - x_{nm})^2+(y_i - y_{nm})^2)/\sigma^2}.
\end{equation}
The location of the CG sites is then obtained by taking the average of this procedure across both leaflets. In this work, we employ $D = 10$ and $\sigma = 15 \ \mathring{\text{A}}$.

\newpage

\subsubsection{Additional Results}
\begin{figure}[H]
        \centering
        \includegraphics[width=1\linewidth]{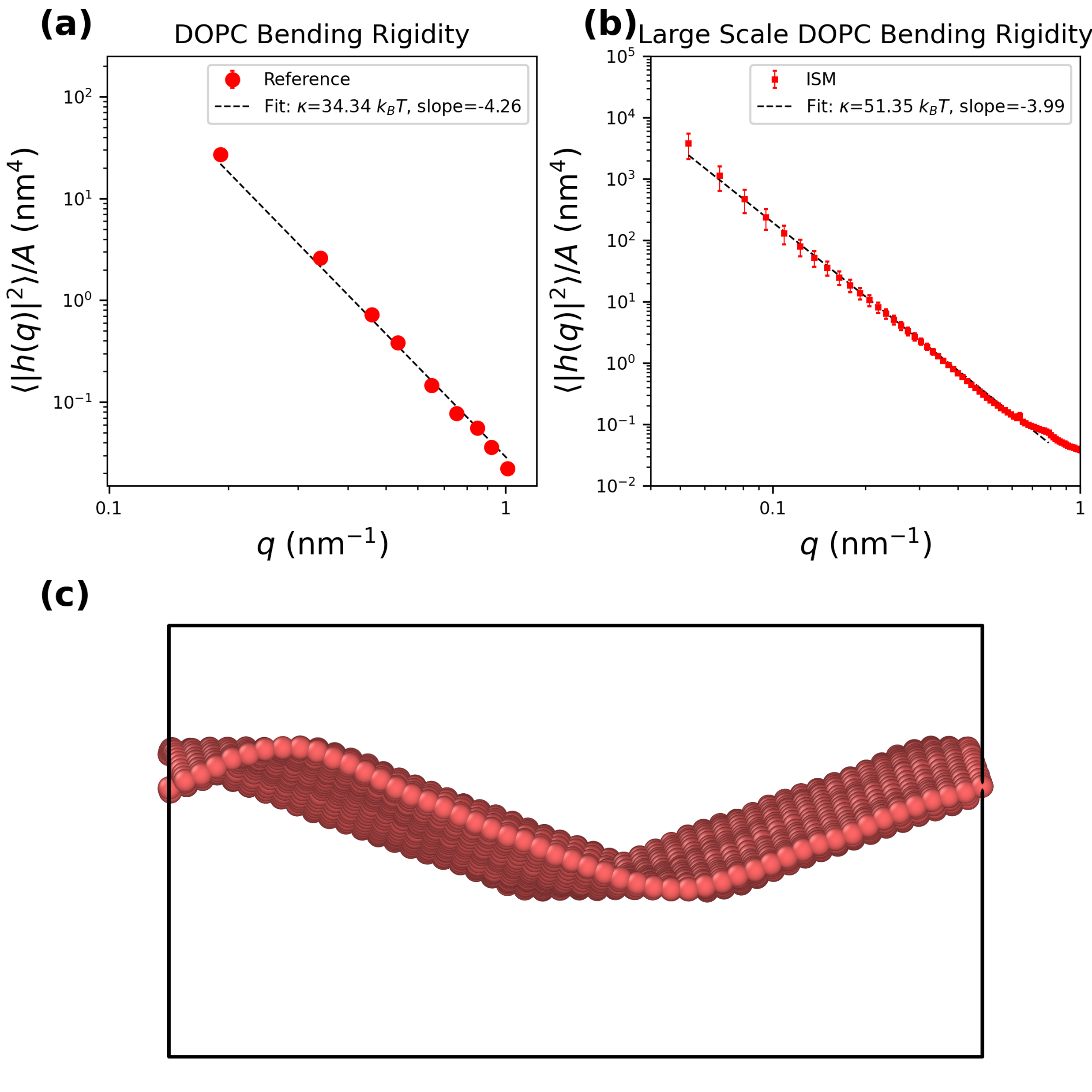}
        \caption{Fluctuation spectrum fits of (a) mapped reference model and (b) ISM model. (c) Buckling of ISM CG DOPC model is shown under lateral compression.}
        \label{fig:DOPC}
\end{figure}

The fluctuation spectrum of the mapped Martini lipid model and the ISM model, as well as the results of the buckling simulation, are shown in Fig. \ref{fig:DOPC}. 
The bending modulus was obtained from fitting of the fluctuation spectrum according to Canham-Helfrich theory, which states that $\langle h^2 \rangle = A/\beta\kappa q^4$, where $\kappa$ is the membrane bending modulus and $A$ is the membrane area.

\subsection{Macroscale Coarse-Graining}
\subsubsection{Renormalization}
Wilsonian renormalization considers the effect of coordinate rescaling, $\mathbf{R} \rightarrow \mathbf{R}/\xi$ on the parameters of a Hamiltonian.
Renormalization was originally developed in the context of fields, which have an infinite number of degrees of freedom.
Furthermore, the terms of the Hamiltonian are determined by the symmetries of the system.
Given the discrete, particle-based nature of our renormalization procedure we make two substitutions here.
The first is that we consider the effect of the rescaling operation on a set of parameters $\phi$ which we denote as the tunable parameters of a full $N$-body expansion, $U[\phi]$, for which any PMF can be fully represented.
To address the finite number of degrees of freedom for our system, we couple our rescaling operation to an addition in particle number by a factor of $\xi^d$ where $d$ is the dimension of the system.
Then, by coarsening the system to the same number of particles $N$ at every operation,  $\boldsymbol{\mathcal{M}}_1 : \mathbf{R}^{\xi\cdot N} \rightarrow \mathbf{R'}^N$, followed by rescaling, $\mathbf{R} = \mathbf{R}'^N/\xi$, this effectively keeps the density of the system constant while enabling rescaling operations to be applied \emph{ad infinitum}.
The renormalization operation $\mathcal{R}$ is then a map $\phi \rightarrow \phi' $ which takes the parameter values of the $N$-body expansion at one length scale to another.
Using the mapping operator defined in Eq., the renormalization operation can be expressed compactly as
\begin{equation}\label{Eq. 29}
    \mathcal{R}[U[\phi]] = - k_BT\cdot \ln \mathbb{E}_{\mathbf{R}^{\xi^d\cdot N}\sim \rho[U[\phi]]}[ \delta(\mathcal{M}_1(\mathbf{R}'^N,\mathbf{R}^{\xi^d \cdot N})/\xi - \mathbf{R}^N)].
\end{equation}
\subsubsection{Additional Results}

When conducting a simulation at a given iteration of the CG procedure, we assume that the self-similarity in the interactions holds such that the same ML model can be used for the PMF at that iteration.
Once a set iteration value is chosen, mass is rescaled according to the average value of $n/N$ at that iteration. 
Denoting the mass of one water molecule as $m_0$, then at $n= 0$ the mass of the CG blobs is $M_0 = 33 * m_0$ for $n/N = 33$, which was the initial CG resolution chosen. 
The iterative CG operations proceeded by multiplying the box lengths by $\xi$ on each dimension while keeping the density constant. 
Consequently, the mass of a CG blob at iteration $n$ is set to be $M_n = \xi^d  * M_0$.
For the CG model at $n = 12$ iterations, this amounts to a blob mass of $8.9 \times 10^{19}$ g/mol, with a single blob constituting on average $4.9 \times 10^{18}$ water molecules.

\section{Additional Information on Machine-Learned Coarse-Grained Simulations}

For each CG water resolution, the CG blob mass was chosen to be the average blob mass.
Droplet simulations proceeded from preformed spherical aggregates. A Lennard-Jones 9-3 interaction was maintained between the $z=0$ surface and CG particles. A value of $\sigma = 1.0$ nm was held constant for all droplet simulations.

Buckling simulations of the CG DOPC model proceeded via simulation of a bilayer patch 25$\times$ the size of the reference bilayer. 
The patch was deformed in the x direction to 0.95 of its original value over 10000 timesteps, which was then proceeded by simulation under the NVT ensemble for another 100000 timesteps.
Vesicle simulations were conducted using a preformed vesicle with a radius of $0.3 \ \mu \text{m}$; CG site number was determined approximately from the ratio of the surface area of the vesicle to CG packing surface area.
The initial CG vesicle configuration was obtained from a spherical Fibonacci lattice packing.
The CG vesicle was then simulated for roughly 170000 timesteps under the NVT ensemble.

Iterative CG models were trained from datasets consisting of 800000 timesteps using a timestep of 2 fs, in which frames were recorded every 1 ps. 
A damping constant of 1 ps was used.
The mass of the CG blobs was set to 1 for iterative CG simulations; 3456 CG blobs were simulated in a box volume which kept the density identical to that of the atomistic model.